\documentclass[aps,prb,floatfix,preprint]{revtex4-2}
\usepackage{amssymb}
\usepackage{amsmath}
\usepackage{graphicx}
\usepackage{float}
\usepackage{caption}

\begin{document}
	
\title{Wide-Mass-Scanning-Range Setup and Phase-Resolved Protocol for Axion Dark Matter Detection}

\author{A. Cottet$^{1,2}$ and T. Kontos$^{1,2,3}$\footnote{To whom correspondence should be addressed: audrey.cottet@espci.fr;takis.kontos@ens.fr}}

\affiliation{Laboratoire de Physique de l'\'{E}cole Normale Sup\'{e}rieure, ENS, Universit\'{e} PSL, CNRS, Sorbonne Universit\'{e}, Universit\'{e} Paris-Diderot, Sorbonne Paris Cit\'{e}, Paris, France.}
\affiliation{Laboratoire de Physique et d'Etude des Mat\'eriaux, ESPCI Paris, PSL University, CNRS, Sorbonne Universit\'e, Paris, France}
\affiliation{Institute of Astrophysics, FORTH, GR-71110 Heraklion, Greece}

\begin{abstract}
We propose a paradigm for quantum enhanced axion dark matter search, which does not rely on power measurements. We propose to measure directly the axion amplitude and phase in an interferometric protocol at the quantum limit, using a non-linear cavity. In addition, we introduce gyromagnetic modes as wide mass range transducers for axion signals compatible with standard haloscope designs. We expect this scheme to offer an improvement of at least 4 orders of magnitude in figure of merit and at least 2 orders of magnitude in mass window with respect to standard haloscopes. Owing to its generality, our proposed protocol has the potential to speed up axion search but also the search for dark photons or other cosmological objects, such as galactic masers.
\end{abstract}

\maketitle

\textbf{Introduction}

There is a general consensus that a large part of the matter and energy in the Universe is unknown. Well established candidates for dark matter are axions or axion-like particles. While axions are supposed, if they exist, to
be everywhere in the galactic halo, their interaction with standard model particles is expected to be very weak and their mass is unknown. Hence, their detection requires broadband and ultrasensitive amplification and measurement techniques. Quantum sensing is appealing because it reaches the ultimate resolution, limited by the Heisenberg uncertainty principle. Pioneering experiments using quantum limited microwave amplifiers based on superconducting circuits technology have explored the possibility to
accelerate the axion dark-matter search. However, most of these methods intrinsically deal with power measurements, which leads to a major limitation in the detection time. Furthermore, the range of accessible frequencies is limited in standard microwave designs.

The basic principle for axion detection was proposed by P. Sikivie in 1983 \cite{Sikivie:83}. It consists of a microwave cavity immersed in a large static magnetic field, called a haloscope. The idea is to reveal the effect of the potential existence of axions \cite{PecceiQuinn:77,Wilczek:78, Weinberg:78,review:1,review:2} on electrodynamics, as a term
\begin{equation}
\hat{H}_{ax}=\frac{1}{\mu _{0}c}g_{a\gamma \gamma }\cos (\omega
_{ax}t-\varphi _{ax})\int d^{3}\vec{r}~\vec{E}.\vec{B}  \label{HcBis}
\end{equation}%
which is expected from cosmological theories\cite{review:1,review:2}. This
hypothetical term in the Hamiltonian of the electromagnetic field involves a
coupling strength $g_{a\gamma \gamma }$ which depends on the axion mass and
density (assumed homogeneous here at the scale of the detector), the axion frequency $\omega _{ax}/2\pi $ which corresponds to the
axion mass, the phase $\varphi _{ax}$ of the axion signal and the electric
and magnetic fields $\overrightarrow{E}$ and $\overrightarrow{B}$. If the
axion frequency is resonant with the cavity microwave resonance frequency $%
\omega _{cav}/2\pi $, axions are expected to produce an excess of photons in
the haloscope, leading to an excess of power leaking out of the detector.

Since the first proposal of Sikivie, pioneering works have pushed
amplification techniques to the required accuracy \cite{Brubaker:17}. This
can be recast in the figure of merit of haloscopes $F_{halo}$, a rate which
corresponds to the inverse of the time to get a signal-to-noise ratio of 1.
Recently, quantum amplification techniques have been used to boost $F_{halo}$
\cite{Malnou:19,Crescini:20,Brubaker:17,Braine:20}. In particular, recent
implementations have used quadrature squeezing and single-photon detection
techniques\cite{Backes:21,QUAX:24}. However, all the haloscopes implemented
so far have focused on power sensing, which is a major hurdle to increase $%
F_{halo}$. With the devices implemented so far, the axion-induced power
leaking out of the detector is expected to be very small, typically of the
order of $10^{-23}W$ \cite{Backes:21} while $F_{halo}$ is limited to the $%
mHz $ range. Furthermore, the standard microwave cavities used in these setups have a limited frequency tunability. This has three major consequences. First, this strongly limits the scanning range of frequencies corresponding to different axion masses.
Second, one can only test the most optimistic scenarios where the density of
axions would be high at the Earth position in the Milky Way \cite{voids:Igor}. Third, this only allows one to probe average quantities for the slowest
haloscopes. However, important time dependent aspects such as daily
variations of axion signals are expected to occur on hour timescales\cite{Turner:90}. In addition, the dephasing dynamics of the axion field is
expected to occur on millisecond timescales \cite{CAPP:22,Lentz:2017}.

Here, we propose to speed up the axion detection via the direct sensing of the axion amplitude and phase. We also propose to use a gyromagnetic mode, which is naturally
magnetic field-tunable, to be able to scan a mass range of several tens of $\mu eV$ with the same detector.  We expect a figure of merit
exceeding by more than four orders of magnitude that of existing detectors.
This method opens the way to real-time detection of hypothetical axion
signals and axion dynamics on sub-millisecond time scales.

\textbf{Principle of the phase-resolved haloscope\label{principle}}

The goal of a haloscope is to measure the effect of the axion term $\hat{H}_{ax}$ of Eq.(\ref{HcBis}) on the electromagnetic field of a microwave
cavity. Our proposed detection scheme is based on two main ideas (see Fig. 1c). First, we
exploit specific chiral modes of a magnetic crystal inserted in the
microwave cavity: the gyromagnetic modes, also called gyrotropic magnetic modes\cite{Gurevich:96}. These modes have the properties of being magnetic field
tunable over several orders of magnitude and being coupled to the axions through the term $\hat{H}_{ax}$ of Eq.(\ref{HcBis}). We will show below that
this has important consequences on the axions frequency range accessible
with our setup. Second, we use a number-number coupling between a "readout"
mode of the microwave cavity and the gyromagnetic mode in order to convert
the axion-photon mixing term $\hat{H}_{ax}$, not into a power, but into a
frequency shift (see Fig. 1b). Such a nonlinear coupling is possible if the microwave
cavity is coupled to a non-dissipative nonlinear element such as a Josephson junction, ideally magnetic field resilient\cite{Lachance:2017}. One can for instance use a granular aluminium Josephson junction\cite{Thery:24,Thery:24b}. The resulting
axion-induced frequency shift of the readout mode leads to a phase shift;
hence the name 'phase-resolved haloscope'. Our detection protocol exploits
this axion-induced phase shift, represented in Figure 1b using the phase
space representation of the electromagnetic field of the readout mode. In
contrast, the conventional haloscope scheme, represented on Figure 1a, involves
an axion-induced increase of the cavity photonic power. We will show
below that, due to this geometric difference in the electromagnetic phase
space, our phase-resolved detection scheme offers major advantages in terms
of the figure of merit. Throughout this work, we set $\hslash =1$ unless
explicitly stated.

We apply the Black-Box Quantization method to our haloscope to write its
Hamiltonian as (see details in the Supplementary Information (SI))\cite{Nigg:2012}:
\begin{eqnarray}
\hat{H} &=&\tilde{\omega}_{a}\hat{a}^{\dag }\hat{a}+\tilde{\omega}_{m}\hat{m}^{\dag }\hat{m}+K_{a}\hat{a}^{\dag 2}\hat{a}^{2}+K_{am}\hat{a}^{\dag }\hat{a}%
\hat{m}^{\dag }\hat{m}  \notag \\
&&+\varepsilon _{a}e^{-i(\omega _{1}t-\varphi _{1})}\hat{a}^{\dag
}+\varepsilon _{a}e^{i(\omega _{1}t-\varphi _{1})}\hat{a}  \notag \\
&&+\varepsilon _{m}e^{-i(\omega _{2}t-\varphi _{2})}\hat{m}^{\dag
}+\varepsilon _{m}e^{i(\omega _{2}t-\varphi _{2})}\hat{m}  \notag \\
&&+\hat{H}_{bath}+\hat{H}_{ax}  \label{1}
\end{eqnarray}
The above equation involves the annihilation operators $\hat{a}$ and $\hat{m}
$ of the readout mode and the gyromagnetic mode. In practice, the
nonlinearity of the microwave circuit generates a number-number coupling
term in $K_{am}$ between the readout and gyromagnetic modes, but also a self
Kerr term in $K_{a}$, with $K_{a}<0$ for a cavity coupled to a Josephson
junction. Both terms appear in the first line of Eq.(\ref{1}). Our protocol
requires to drive the readout and gyromagnetic modes, as described by the
real and positive drive amplitudes $\varepsilon _{a}$ and $\varepsilon _{m}$. The term in $\hat{H}_{bath}$ describes the baths coupled to the readout
and gyromagnetic modes, which result in finite linewidths $\Lambda _{a}$ and
$\Lambda _{m}$. Our detection scheme is based on a conversion of axions into
readout photons, mediated by the gyromagnetic mode, as sketched in Figure 1d. This effect is enabled
by the axion contribution of Eq.(\ref{HcBis}), which appears in the full
haloscope Hamiltonian of Eq.(\ref{1}). As we derive in the next section and
in the SI, $\hat{H}_{ax}$ can be reexpressed as:
\begin{equation}
\hat{H}_{ax}=[\Omega _{ax}^{0}(\hat{a}+\hat{a}^{\dag })+\Omega _{ax}(\hat{m}+\hat{m}^{\dag })]\cos (\omega _{ax}t-\varphi _{ax})  \label{2}
\end{equation}
The term in $\Omega _{ax}^{0}$ in (\ref{2}) is the term used in the Sikivie haloscope whereas the term in $\Omega _{ax}$ is the one arising naturally from the coupling of the gyromagnetic mode to the axion field (see below Eq. (7)). Both terms are proportional to the
axion photon coupling constant $g_{a\gamma \gamma }$ of Eq.(\ref{HcBis}).
They both contain a geometrical factor $\int d^{3}\vec{r}~\vec{E}.\vec{B}$
which contributes to the "form factor" from the axion detection literature. Anticipating on the specific design discussion in the section devoted to the gyromagnetic mode, we focus on the $TM_{110}$ mode in a cubic cavity (z direction along the external magnetic field) which is equivalent to the $TM_{010}$ mode in cylindrical cavities used in many conventional haloscope designs. This ensures a large form factor as needed for axion detection.
However, the effective value of $\int d^{3}\vec{r}~\vec{E}.\vec{B}$ is
different for $\Omega _{ax}^{0}$ and $\Omega _{ax}$ because $\Omega_{ax}^{0} $ involves a direct contribution from the electric field
associated to the readout mode whereas $\Omega _{ax}$ involves the electric
field associated to the gyromagnetic mode. Note that in conventional
haloscopes, the role of the static magnetic field $\overrightarrow{H_{0}}$
is to induce $\Omega _{ax}^{0}\neq 0$ thanks to the coupling between the
magnetic induction $\overrightarrow{B_{0}}=\mu _{0}\overrightarrow{H_{0}}$
and the cavity microwave electric field in Eq.(\ref{HcBis}). In our
phase-resolved haloscope, a finite magnetic field helps to increase $\Omega_{ax}$, which is already finite for $H_{0}=0$ due to the DC magnetization $\overrightarrow{M_{0}}$ of the magnetic crystal. Furthermore, the DC
magnetic field has another major role, which is to tune the frequency $\tilde{\omega}_{m}$ of the gyromagnetic mode. Eventually, as we already
explained above in plain language, our protocol is sensitive to the
phase $\arg [\langle \hat{a}\rangle ]$ whereas all conventional haloscopes
focus on the power and therefore the expectation value of $\langle \hat{a}^{\dag }\hat{a}\rangle $.

The general principle of our detection scheme is as follows. The gyromagnetic mode is simultaneously driven by the
pump drive with amplitude $\varepsilon _{m}$ and the axion-induced drive with amplitude $\Omega _{ax}$, at the
semiclassical level. These terms lead to an amplitude for the $\hat{m}$ field which is essentially the sum of two terms: one arising from the $\varepsilon _{m}$ drive which we note $\sqrt{n_{m}}e^{i(\omega _{2}t+\phi_{2})}$ and one arising from the axion amplitude $\sqrt{n_{ax}}e^{i(\omega_{ax}t+\varphi _{ax})}$. These two amplitudes can interfere constructively
or destructively  which modifies the
frequency of the readout mode through the nonlinear term in $K_{am}$. More precisely, the $K_{am}$ cross-Kerr term
leads to a cavity readout frequency shift $\Delta \omega _{a}=K_{am}\langle
\hat{m}^{\dag }\hat{m}\rangle \approx K_{am}|\sqrt{n_{ax}}e^{i\varphi _{ax}}+\sqrt{n_{m}}e^{i\varphi _{2}}|^{2}$ if $\omega _{2}=\omega _{ax}=\omega _{m}$
. It contains an interference term, $\Delta \omega _{a}^{cross}=2K_{am}\sqrt{n_{m}}\sqrt{n_{ax}}\cos (\varphi _{ax}-\varphi _{2})$, which corresponds to
the signal in our detection scheme. The fact that this term can always be
made resonant by changing $\omega _{2}$ and $\omega _{m}$ is what gives rise
to the tunable character of our detection scheme. Besides, the very fact
that $\Delta \omega _{a}^{cross}$ is a frequency shift directly reflects the
phase-resolved nature of our protocol. With such a scheme, a large number of
photons in the cavity, induced by a large value of $\varepsilon _{a}$, can
be used to boost the axion signal, since it will enable one to convert $\Delta \omega _{a}^{cross}$ into a large phase shift. We will discuss
below the choice of the optimal value for $\varepsilon _{a}$, which is not
trivial due to the self Kerr term in $K_{a}$.
Interestingly, the phase-resolved character of our protocol is twofold
since it relies on the phase shift measurement of the readout mode and it
also conveys information on the phase of the axion field, as visible from
the factor $\cos (\varphi _{ax}-\varphi _{2})$ in $\Delta \omega
_{a}^{cross} $. Finally, the expression of $\Delta \omega _{a}^{cross}$
reveals that the axion signal can be amplified not only by a large $\varepsilon _{a}$ but also by a strong number of excitations in the
gyromagnetic mode, and it is therefore interesting to use a sufficiently
strong value for $\varepsilon _{m}$.

Evaluating the figure of merit of a detector requires estimating its signal
but also its noise. The phase-resolved nature of our detector has profound
consequences on its figure of merit, as we will see below. This provides
another major advantage of our scheme in comparison with conventional
haloscopes or haloscopes based on single-photon detection \cite{Dixit:19,QUAX:24}. However, as in conventional haloscopes, the expected
signal is much smaller than the quantum fluctuations of $\hat{a}$. It is
thus necessary to take into account fully the quantum noise of the fields $%
\hat{a}$ and $\hat{m}$. Such a full quantum description of the detector can
be conveniently carried out using the Keldysh path integral technique, which
is well suited for incorporating dissipation effects in $\Lambda _{a}$ and $\Lambda _{m}$, induced by the baths coupled to the haloscope\ \cite{Cottet:20}.

\begin{figure}[tph]
\begin{center}
{\small \includegraphics[width=0.55\linewidth,angle=0]{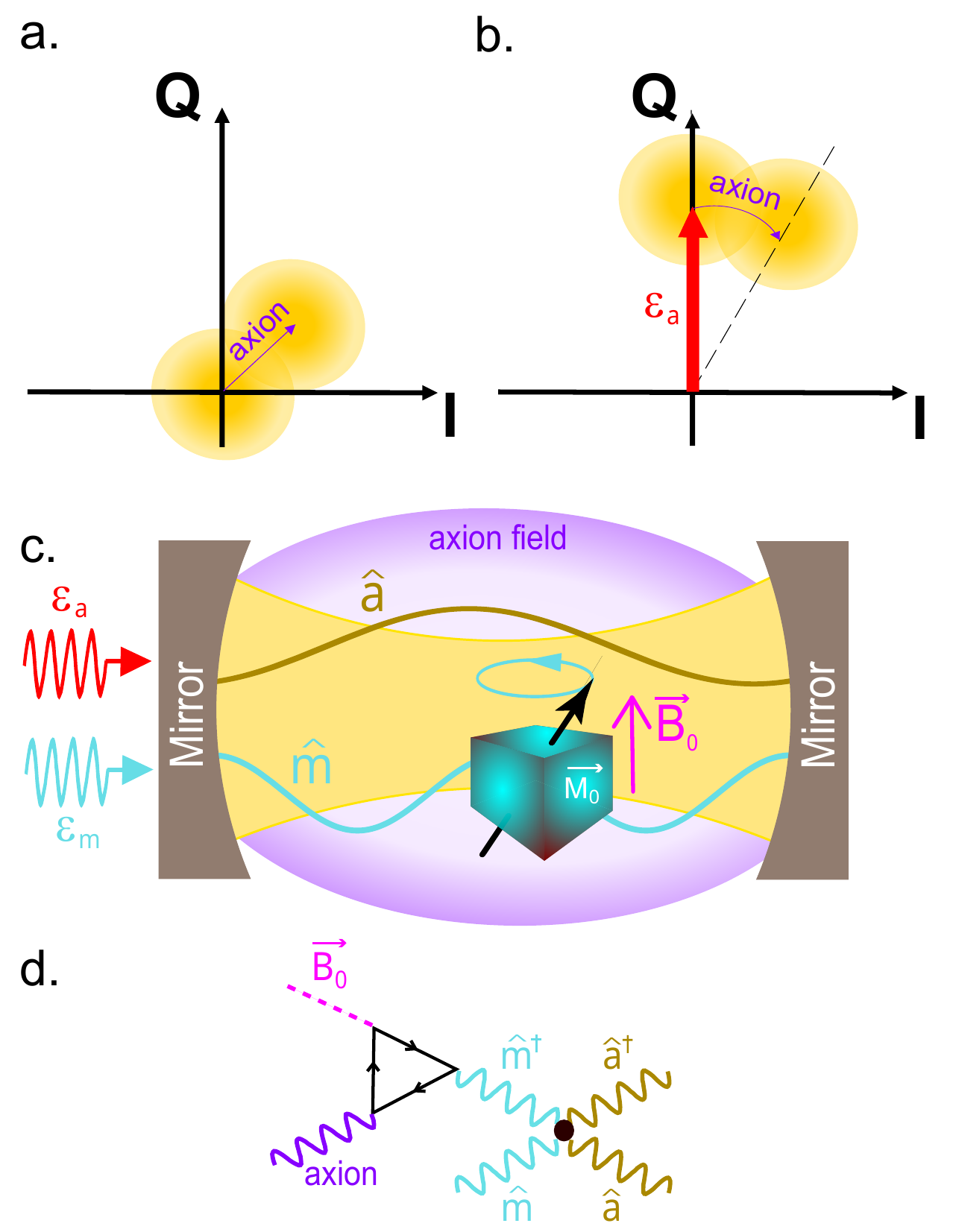}} \caption{\textbf{Principle of the phase-resolved haloscope}\textbf{(a)}: Conventional haloscope concept represented in the I-Q (quadratures) plane of the cavity electromagnetic field. The axion field displaces the
cavity field away from the center of the I-Q plane (vacuum). This phase can take any value in the I-Q plane with no
consequence on the detection process. \textbf{(b)}: Phase-resolved haloscope. The cavity drive with amplitude $\varepsilon_a$ displaces the cavity field and the axion changes its phase. \textbf{(c)}: Architecture of our phase-resolved haloscope. The magneto-electric coupling results from the presence of the axion field. The axion field in purple drives the gyromagnetic mode (in light blue) thanks to the magnetic field $\overrightarrow{B_{0}}$ and the magnetization $\overrightarrow{M_{0}}$ of the gyromagnetic crystal in magenta. This signal is then converted into a signal in the readout mode (in brown) thanks to the nonlinear coupling in $K_{am}$. The magnetic field tunability of the
gyromagnetic frequency $\omega _{m}$ enables the scanning property of the phase-resolved haloscope.\textbf{(d)}: Diagrammatic representation of the axion-magnetic mode conversion process (triangular vertex) and its read-out through the cross-Kerr number-number point interaction (black dot).}
\label{fig:principle}
\end{center}
\end{figure}

\textbf{Axion-photon coupling in the presence of a gyromagnetic mode\label{CouplingValue}}

In this section, we evaluate the coupling between the gyromagnetic mode and
the hypothetical axion field, which appears in  Eq.~(\ref{HcBis}). We assume that
the gyromagnetic mode is hosted by a magnetic chiral crystal such as YIG,
inserted into a microwave cavity~\cite{Tobar:19}. We describe the dynamics
of the magnetization $\vec{M}$ of the gyromagnetic crystal with a
Landau-Lifshitz-Gilbert (LLG) equation in order to derive the permeability
matrix of the magnetic material. We define the external magnetic field $\vec{H}_{0}=H_{0}\vec{z}$ and the static magnetization $\vec{M}_{0}=M_{0}\vec{z}$. The AC permeability matrix associated to the gyromagnetic mode reads, in a
basis $\{\vec{x},\vec{y},\vec{z}\}$~\cite{Gurevich:96}:

\begin{equation}
\hat{\mu}(\omega) = \mu_0 \left[
\begin{array}{ccc}
\mu_r(\omega) & i\mu_a(\omega) & 0 \\
-i\mu_a(\omega) & \mu_r(\omega) & 0 \\
0 & 0 & 1%
\end{array}
\right],  \label{acmpBis}
\end{equation}

The gyromagnetic modes are associated to an electric field polarized along
the magnetization $\vec{M}_{0}$ and constant along that direction. We
disregard magnetic dissipation, since we are simply seeking for the energy of the gyromagnetic modes. Importantly, the elements $\mu _{r}$ and $\mu _{a}$ of the permeability matrix $\hat{\mu}$ are strongly magnetic-field
dependent. The boundary conditions for a box-shaped gyromagnetic crystal
with volume $V=L_{x}L_{y}L_{z}$ (as sketched in Fig.~1c) yield
eigenfrequencies $\omega _{n,l,s}$ indexed with integers $n$, $l$ (see
details in the SI); the exact shape  of the gyromagnetic crystal is unimportant). The index $s=\pm 1$ corresponds to two solution
branches. The frequencies $\omega _{n,l,s}$ are naturally tunable with the
magnetic field $H_{0}$ and the DC magnetization $M_{0}$, via the parameters $%
\mu _{a}(\omega )$ and $\mu _{r}(\omega )$. Importantly, owing to the axial
symmetry of the electric field for these gyromagnetic modes, many of them
correspond to a non-zero geometrical factor $\int d^{3}\vec{r}\,\vec{E}\cdot
(\mu _{0}\vec{H}_{0}+\vec{M}_{0})$.

The standard field quantization procedure can be applied to the gyromagnetic
modes. Below, we explicitly write $\hbar $ in the equations. The
corresponding contribution to the Hamiltonian is (see SI):%
\begin{equation}
\hat{H}=\sum_{n,l,s}\hbar \omega _{n,l,s}\left( \hat{m}_{n,l,s}^{\dag }\hat{m}_{n,l,s}+\frac{1}{2}\right)
\end{equation}
The general axion-photon coupling of Eq.(\ref{HcBis}) writes, for the
gyromagnetic modes considered here (see Eqs.(99) and (100) of
SI):
\begin{equation}
\hat{H}_{ax}=\sum_{\substack{ n,l\text{ odd}  \\ s\in \{+1,-1\}}}\hbar
\Omega _{ax}^{n,l,s}\cos (\omega _{ax}t-\varphi _{ax})(\hat{m}_{n,l}+\hat{m}_{n,l}^{\dag })  \label{driveBis}
\end{equation}
with
\begin{equation}
\Omega _{ax}^{n,l,s}=\frac{1}{\hbar \mu _{0}c}g_{a\gamma \gamma }\mu_{0}(M_{0}+H_{0})\frac{4}{ \pi ^{2}nl}\sqrt{\frac{V}{2\varepsilon }\hbar \omega _{n,l}}
\label{omega_ax_nous}
\end{equation}
where $n$ and $l$ are odd integers and $\varepsilon $ is the permittivity of
the magnetic material (\textit{for even integers, the axion does not couple to the gyromagnetic mode}). The prefactor $\frac{4}{ \pi ^{2} nl}$ is proportionnal to the square root of the 'form factor' $C^{n,l}$ of the (n,l,0) mode found in the axion litterature. Equation (\ref{omega_ax_nous}) shows that the value
of the coupling between the axion field and the gyromagnetic modes is finite
even when $H_{0}=0$, and it can be increased using a finite $H_{0}$. Below, we focus on the mode which has the strongest geometrical factor $\int d^{3}\vec{r}~\vec{E}.\vec{B}$ in the box-shaped geometry, which is for $n=l=1$
from Eq.(\ref{omega_ax_nous}). Furthermore, we focus on the branch $s=+1$ of
this mode, which shows the largest frequency dispersion. We thus define for
the rest of the paper: $\omega _{m}=\omega _{1,1,+1}$ and $\Omega
_{ax}=\Omega _{ax}^{1,1,+1}$. To evaluate the values of $\omega _{m}$ and $\Omega _{ax}$, we use the realistic parameters of Table S1 in the SI corresponding to a YIG crystal. We find that when the field $\mu _{0}H_{0}$
varies between $0~$\textrm{T} and $2~$\textrm{T, the frequency }$\omega _{m}$
varies increases from $6.9~\mathrm{GHz}$ to $63~\mathrm{GHz}$ and the
coupling $\Omega _{ax}$ varies from $-17Hz$ to $-463~$\textrm{Hz} (see Fig.
S1 in the SI). A coupling $\left\vert \Omega _{ax}\right\vert \geq 200~$\textrm{Hz} is obtained for $\omega _{m}\in \lbrack 36~\mathrm{GHz,}63~\mathrm{GHz}]$. We thus obtain a strong variation in $\omega _{m}$, along with a sizeable value for $\Omega _{ax}$. A particularly convenient
parameter to adjust the range covered by $\Omega _{ax}$ is $L_{z}$, which
modifies $\Omega _{ax}$ but not $\omega _{m}$.

\textbf{Effective description of the readout cavity after integrating out
the gyromagnetic mode\label{EffectiveDescription}}

In order to predict the signal in the readout mode, it is convenient to
integrate out the gyromagnetic mode from the haloscope description
corresponding to the Hamiltonian of Eq.(\ref{1}). The Keldysh path integral
method is particularly suitable for this task, because it accounts for the
dissipation sources inherent to the readout and the gyromagnetic modes, and
it enables a systematic treatment of the haloscope signal statistics \cite{Kamenev:2011,Sieberer:2016}. We therefore use it here to derive the
effective dynamics of the readout mode in the presence of the gyromagnetic
mode and the axion mode. The action of the coupled readout mode,
gyromagnetic mode and axion field reads\cite{Cottet:20}:
\begin{align}
S_{tot}& =\int_{-\infty }^{+\infty }dt\left( [\bar{\bar{\varphi}}_{a,cl}(i\partial
_{t}-\omega _{a}-i\Lambda _{a}/2)\varphi _{a,q}-K_{a}(\left\vert \varphi
_{a,cl}\right\vert ^{2}+\left\vert \varphi _{a,q}\right\vert ^{2})\bar{%
\bar{\varphi}}_{a,cl}\varphi _{a,q}\right. \notag \\
& -\sqrt{2}(\varepsilon _{a1}e^{-i(\omega _{1}t-\varphi _{1})}+\Omega
_{ax}^{0}\cos (\omega _{ax}t-\varphi _{ax}))\bar{\bar{\varphi}}_{a,q} \notag \\
& +\bar{\bar{\varphi}}_{m,cl}(i\partial _{t}-\omega _{m}-i\Lambda _{m}/2)\varphi
_{m,q}-\sqrt{2}(\varepsilon _{m}e^{-i(\omega _{2}t-\varphi _{2})}+\Omega
_{ax}\cos (\omega _{ax}t-\varphi _{ax}))\bar{\bar{\varphi}}_{m,q}  \notag \\
& -\frac{K_{am}}{2}[(\left\vert \varphi _{a,cl}\right\vert ^{2}+\left\vert
\varphi _{a,q}\right\vert ^{2})\bar{\bar{\varphi}}_{m,cl}\varphi
_{m,q}+(\left\vert \varphi _{m,cl}\right\vert ^{2}+\left\vert \varphi
_{m,q}\right\vert ^{2})\bar{\bar{\varphi}}_{a,cl}\varphi _{a,q}])]+h.c.  \notag \\
& \left. +i\Lambda _{m}(2n_{B}(\omega _{m})+1)\bar{\bar{\varphi}}_{m,q}\varphi
_{m,q}+i\Lambda _{0}(2n_{B}(\omega _{a})+1)\bar{\bar{\varphi}}_{a,q}\varphi
_{a,q}\right) \label{fullaction}
\end{align}

The fields $\varphi _{a,cl/q}(t)$ and $\varphi _{m,cl/q}(t)$
in the above equation are the classical/quantum parts of the variables
associated to the readout mode and the gyromagnetic mode on the Keldysh time contour. We note $\bar{\bar{\varphi}}$ the complex conjugate of any field $\varphi $. We have included the dissipation rates $\Lambda _{a}$ and $\Lambda _{m}$ for the readout and gyromagnetic modes in the Markovian
approximation (see details in the SI). The above action has the
property of being quadratic in the variables of the gyromagnetic mode. One
can therefore integrate out exactly the gyromagnetic mode to obtain the
effective action for the readout mode. At second order in $K_{am}$, this
effective action takes the form

\begin{align}
S_{a,eff}& =\int_{-\infty }^{+\infty }dt\bar{\bar{\varphi}}_{a,cl}(i\partial _{t}-\breve{\omega}_{a}-i\Lambda _{a}/2)\varphi _{a,q}-\widetilde{K}_{a}(\left\vert \varphi
_{a,cl}\right\vert ^{2}+\left\vert \varphi _{a,q}\right\vert ^{2})\bar{\bar{\varphi}}_{a,cl}\varphi _{a,q}-\sqrt{2}\breve{\varepsilon}(t)\bar{\bar{\varphi}}%
_{a,q}+h.c \label{EffectiveAction} \\
& +i\Lambda _{a}(2n_{B}(\omega _{a})+1)\bar{\bar{\varphi}}_{a,q}\varphi _{a,q}+i K_{am}^2 \frac{\Lambda _{m}}{2} (2n_{B}(\omega _{m})+1) \frac{\varepsilon^{2}_{m}}{((\omega
_{2}-\omega _{m})^{2}+\frac{\Lambda _{m}^{2}}{4})^2}[ \bar{\bar{\varphi}}_{a,cl}\varphi _{a,q}+\bar{\bar{\varphi}}_{a,q}\varphi _{a,cl} ]^2
\notag
\end{align}

with
\begin{equation}
\breve{\varepsilon}(t)=\varepsilon _{a}e^{-i(\omega _{1}t-\varphi
_{1})}+\Omega _{ax}^{0}\cos (\omega _{ax}t-\varphi _{ax})  \label{epstild}
\end{equation}

\begin{equation}
\breve{\omega}_{a}=\tilde{\omega}_{a}+\frac{1}{2}\frac{\Omega
_{ax}K_{am}\varepsilon _{m}e^{i(\omega _{2}-\omega _{ax})t+\varphi
_{ax}-\varphi _{2})}}{(\omega _{2}-\omega _{m}-i\Lambda _{m}/2)(\omega
_{ax}-\omega _{m}+i\Lambda _{m}/2)}+h.c  \label{omegaadresssed}
\end{equation}%
\begin{equation}
\tilde{\omega}_{a}=\omega _{a}+K_{am}(\frac{1}{2}+n_{B}(\omega _{m}))+\frac{%
K_{am}\varepsilon _{m}^{2}}{((\omega _{2}-\omega _{m})^{2}+\frac{\Lambda
_{m}^{2}}{4})}  \label{watild}
\end{equation}%
and%
\begin{equation}
\widetilde{K}_{a}=K_{a}+K_{am}^{2}(\omega _{2}-\omega _{m})\frac{%
\varepsilon _{m}^{2}}{((\omega _{2}-\omega _{m})^{2}+\frac{\Lambda _{m}^{2}}{%
4})^{2}}.  \label{Katild}
\end{equation}%
The saddle point equation of the effective action of Eq.(\ref{EffectiveAction}) allows us to determine the semiclassical signal of the
haloscope whereas the full partition function associated to $S_{tot}$
entails the noise of the signal\cite{Cottet:20}. Importantly, from Eq.(\ref%
{omegaadresssed}), the dressed readout mode frequency $\breve{\omega}_{a}$
contains a parametric mixing term proportional to $\Omega _{ax}K_{am}$,
which shifts the cavity photon frequency due to the interference between the
gyromagnetic mode and the axion mode. This contribution corresponds exactly
to the term heuristically derived below Equation (3). From Eq.(\ref{watild}), the readout mode frequency also contains a dispersive shift term in $K_{am}$
directly caused by the gyromagnetic mode, similar to what has recently been
observed for a cavity coupled to magnon modes \cite{Lachance-Quirion:17}. As
expected, the effective drive $\breve{\varepsilon}(t)$ applied to the
readout mode contains the term proportional to $\Omega _{ax}^{0}$, which is
the one used in conventional haloscopes. One can already see from the above
Eqs. that the terms in $\Omega _{ax}^{0}$ and $\Omega _{ax}$ do not have the
same effect on the readout mode since the term in $\Omega _{ax}^{0}$
modifies the cavity drive whereas the term in $\Omega _{ax}$ \ directly
modifies the cavity frequency. Importantly, the effect of $\Omega _{ax}^{0}$
is resonant only if the axion, the cavity and cavity drive are resonant,
i.e. $\omega _{a}\approx \omega _{1}\approx \omega _{ax}$. This implies
severe constraints on haloscope axion searches since the axion mass is
unknown and standard microwave cavity have a poorly tunable $\omega _{a}$.
Interestingly, we relax this constraint here since the new term in $\Omega
_{ax}$ can always be made resonant with the axions by properly tuning the
gyromagnetic mode frequency with $B_{0}$ such that $\omega _{m}\approx
\omega _{2}\approx \omega _{ax}$. As illustrated by the previous section,
with gyromagnetic crystals, $\omega _{m}$ can be tuned with a dc magnetic
field on a range of several GHz or tens of GHz\cite{Gurevich:96,Thery:24b}. It is interesting to note that such a feature as well as a magnetic field resilience of a superconducting circuit allowing to readout gyromagnetic modes up to $2.4T$ has been realized experimentally recently \cite{Thery:24b}.
One can even imagine to use several gyromagnetic modes to span an even larger frequency
interval with the same detector.

To give an intuitive understanding of the effective cavity action of Eq.(\ref{EffectiveAction}), it is useful to rewrite it in the form of an effective
cavity Hamiltonian. For simplicity, we consider the resonant condition $\omega
_{2}=\omega _{ax}=\omega _{m}$. One can define photon numbers $%
n_{m}=(2\varepsilon _{m}/\Lambda _{m})^{2}$ and $n_{ax}=(\Omega
_{ax}/\Lambda _{m})^{2}$ induced by the drives in $\varepsilon _{m}$ and $%
\Omega _{ax}$ in the linear regime ($K_{a}=0$ and $K_{am}=0$), with the factor of 2 in the first definition only because, in the Hamiltonian (\ref{1}), $\varepsilon _{m}$ multiplies an exponential term $e^{-i(\omega_{2}t-\varphi _{2})}$ whereas $\Omega _{ax}$ multiplies the cosine term $%
\cos (\omega _{ax}t-\varphi _{ax})$. In this case, one can check that the
action $S_{a,eff}$ maps onto the Hamiltonian:%
\begin{eqnarray}
\hat{H}_{a,eff} &=&(\tilde{\omega}_{a}+2K_{am}\sqrt{n_{ax}}\sqrt{n_{m}}\cos
(\phi _{ax}-\phi _{2}))\hat{a}^{\dag }\hat{a}  \label{Heffheur} \\
&&+K_{a}\hat{a}^{\dag 2}\hat{a}^{2}+\breve{\varepsilon}(t)\hat{a}%
^{\dag }+\breve{\varepsilon}(t)^{\ast }\hat{a}+\hat{H}_{bath,a}^{eff}  \notag
\end{eqnarray}
with $\breve{\varepsilon}(t)$ and $\tilde{\omega}_{a}$ given by Eqs.(\ref{epstild}) and (\ref{watild}). The above
expression makes a direct contact between the full derivation of the
effective action of our system and the heuristic argument  presented previously. It shows again that the axion signal directly renormalizes the readout mode
frequency. It also shows in a compact way the fundamental difference of the
phase-resolved haloscope and the other schemes measuring a photonic power
(standard haloscopes or single photon detectors), which do not involve the
same term of the readout mode effective Hamiltonian. In conventional setups,
the axion field contributes to the cavity drive term $\breve{\varepsilon}(t)$%
. The term $\hat{H}_{bath,a}^{eff}$ describes the net dissipation which
affects the cavity. It accounts for the $\Lambda _{a}$ line broadening which
originally affects the cavity, plus a new dephasing rate
\begin{equation}
\Lambda _{\phi }=\frac{16K_{am}^{2}\varepsilon _{m}^{2}}{\Lambda _{m}^{3}}%
(1+2n_{B}(\omega _{m}))  \label{LandaPhi}
\end{equation}%
which derives from the term proportional to $K_{am}^{2}\varepsilon _{m}^{2}$
in Eq.(\ref{EffectiveAction}) (see details in the SI section III.3). This term is
directly linked to number fluctuations in the driven mode $\hat{m}$,
equivalent to the shot noise of the coherent state drive. It leads to a
diffusion of the cavity field in the I-Q plane of Fig.1b, which blurs the
visibility of the axion-induced phase shift. This term has no classical
equivalent and drops out at the semiclassical level. It imposes a limit on
the figure of merit of the phase-resolved haloscope, as discussed at the end of the main text\ref{choiceEpsm}. Anticipating on the working point chosen for our setup, we
will make sure that our proposed set of parameters is in a regime where $\Lambda _{\phi }\ll \Lambda _{a}$. Hence, unless specified, we now neglect
the $\Lambda _{\phi }$ term. In the following, we will work with the
effective action of Eq.(\ref{EffectiveAction}) instead of the Hamiltonian (\ref{Heffheur}) because we need to consider on and off resonant conditions.
We will disregard the term in $\Omega _{ax}^{0}$ except  the section I of the SI.

\textbf{Definition of the axion signal and its variance}

The conversion of the axion signal into a frequency shift of $\left\langle
\hat{a}\right\rangle $ is at the heart of our phase-resolved haloscope.
However, the actual measurements require an averaging during a time $T$ \cite{Clerk:10} and a demodulation with a phase $\varphi _{d}$. One can define
the signal $\mathcal{I}$ and its variance $\mathcal{V}$ as:%
\begin{equation}
\mathcal{I}=\int_{0}^{T}dt~\hat{a}_{\Sigma }(t)\cos \left( \omega
_{1}t-\varphi _{d}\right)  \label{Iexpr}
\end{equation}%
\begin{equation}
\mathcal{V}=\iint_{0}^{T}dtdt^{\prime }\left\langle (\hat{a}_{\Sigma }(t)\hat{a}_{\Sigma }(t^{\prime })\right\rangle \cos \left( \omega _{1}t-\varphi
_{d}\right) \cos \left( \omega _{1}t^{\prime }-\varphi _{d}\right)
\label{Vexpr}
\end{equation}
with%
\begin{equation*}
\hat{a}_{\Sigma }(t)=\hat{a}(t)+\hat{a}^{\dag }(t)
\end{equation*}
Note that the field in the readout mode of the cavity is measured indirectly
through a readout line with impedance $Z$. The practical values of the
signal and its variance measured though the readout line can be defined as
\begin{equation}
\mathcal{I}_{out}=\sqrt{Z\hslash \tilde{\omega}_{a}\Lambda _{a}}\mathcal{I/T}
\label{I_out}
\end{equation}%
and
\begin{equation}
\mathcal{V}_{out}=Z\hslash \tilde{\omega}_{a}\Lambda _{a}\mathcal{V/T}^{2}
\label{V_out}
\end{equation}%
Here $\mathcal{I}_{out}$ corresponds to an average external voltage. The
definitions (\ref{I_out}) and (\ref{V_out}), which can be derived with the
standard input-output formalism, enable one to capture correctly the
squeezing of the output field which remains very limited in our protocol\cite{Walls:94}.

\textbf{Choice of parameters\label{parameters}}

Below, we will estimate the figure of merit of our axion detector for the
realistic parameters indicated in Table 1. We note $T_{sys}$ the temperature
of the device. One must check that the renormalized cavity frequency $\tilde{\omega}_{a}$ which appears in Eq.(\ref{omegaadresssed}) remains accessible
with state-of-the-art microwave equipment. With the parameters of Table
1 one obtains an accessible value $\tilde{\omega}_{a}=2\pi \times 6$\textit{~}\textrm{GHz}. We have found $\Omega _{ax}<0$
for the example of a YIG crystal, but the minus sign in $\Omega _{ax}$ can
be incorporated into the value of $\varphi _{ax}^{0}$ and therefore we will
use $\Omega _{ax}>0$. For simplicity, we use a value of $\Omega _{ax}$ which
is independent from $\omega _{m}$, with an order of magnitude compatible
with our quantitative predictions.
\begin{equation*}
\begin{tabular}{|l|l|l|}
\hline
$\omega _{a}/2\pi =5$\textit{~}\textrm{GHz} & $\Lambda _{a}/2\pi =5$\textit{~}\textrm{MHz} & $K_{a}/2\pi =-0.212$\textit{~}\textrm{Hz} \\ \hline
$K_{am}/2\pi =1$\textit{~}\textrm{kHz} & $\varepsilon _{m}=158\Lambda _{m}$
& $\Lambda _{m}/2\pi =1$\textit{~}\textrm{MHz} \\ \hline
$k_{B}T_{sys}=50$\textit{~}\textrm{mK} & $\Omega _{ax}/2\pi =463$\textit{~}%
\textrm{Hz} & $\varkappa _{ax}/2\pi =10^{3}$\textit{~}\textrm{Hz} \\ \hline
\end{tabular}
\end{equation*}

\begin{center}
Table 1. \textbf{Parameters used for the phase resolved haloscope}. This set of parameters is one possible working point yielding a cosmologically relevant sensitivity for our haloscope. The corresponding figure of merit is summarized in Figure 6.
\end{center}

\textbf{Frequency dependence of the axion signal\label{SignalRes2}}

In the limit of a large number of photons in the cavity, the average value $%
\left\langle \hat{a}(t)\right\rangle $ which one needs to insert in the
expression (\ref{Iexpr}) of the signal is very close to the semiclassical
value $a(t)$ of $\hat{a}(t)$ (we will find typically $2\times 10^6$ photons in the
readout mode in the optimal regime, as shown in Fig. S2 of the SI, green curve). The semiclassical value $a(t)$ can be calculated from the saddle point of the
effective cavity action of Eq.(\ref{EffectiveAction}) at first order in $\Omega _{ax}$. The details of this calculation are given in the SI, section IV. It
is necessary to take into account that, in practice, the phase $\varphi_{ax} $ of the axion signal is expected to fluctuate. Since the signal of our detector is a phase shift, it should be sensitive to this effect. The minimal model for the fluctuations of $\varphi _{ax}$ is that is an
inhomogeneous broadening of the velocity distributions in the Galactic halo \cite{Krauss:85,Turner:90,CAPP:22}. We assume that this leads to Gaussian
fluctuations of $\varphi _{ax}(t)$. In this limit, one can write $\langle
e^{i\varphi _{ax}(t)}\rangle \approx e^{i\varphi _{ax}^{0}}e^{-\kappa
_{ax}t} $ if one considers a measurement which starts at time $t=0$ with a
known phase $\varphi _{ax}(t=0)=\varphi _{ax}^{0}$. Assuming that the axion
linewidth $\varkappa _{ax}$ is much smaller than $\Lambda _{a}$, the
equation of evolution for $a(t)$ can be integrated for a constant value of $\varphi _{ax}$ and then the substitution of $e^{i\varphi _{ax}}$ by $e^{i\varphi _{ax}^{0}}e^{-\kappa _{ax}t}$ can be performed in a second step
to get the value of $a(t)$ with a good approximation (see details in
the SI). The full expression of $a(t)$ is rather cumbersome because
many frequencies are involved in the description of our device: the drive
frequencies $\omega _{1}$ and $\omega _{2}$, the gyromagnetic and readout
mode frequencies $\omega _{m}$ and $\tilde{\omega}_{a}$ and the axion
frequency $\omega _{ax}$ . We assume that the frequency drives applied to
the haloscope are constantly tuned in the resonant regime $\omega _{1}=\tilde{\omega}_{a}$ and $\omega _{2}=\omega _{m}$ to get strong signals and
simplify the equations, while $\omega _{2}=\omega _{m}$ is swept in order to
determine $\omega _{ax}$. We choose the detection phase $\varphi _{d}$ which
appears in Eq.(\ref{Iexpr}) such that $\mathcal{I}=0$ in the absence of the
axions. This choice avoids an unnecessary signal background which would
complicate the measurements. This can be achieved using:%
\begin{equation}
\varphi _{d}=\varphi _{1}+\arctan (J_{a})\left[ \pi \right]
\end{equation}%
where the coefficient $J_{a}$ is an analytic function of $\varepsilon _{a}$
(see Eqs. (129) and (130) of the SI for details). Two
different detection phases are possible (modulo $2\pi $), which yield
signals $\mathcal{I=\pm }I_{1}$ with opposite signs but the same modulus. The index $1$ in the notation $I_{1}$ means that we calculate the signal as a first order term in $\Omega
_{ax}$ (see SI, section IV.5).

Figures 2 and 3 show the variations of the signal $I_{1}$ with $\omega
_{m}-\omega _{ax}$ for parameters indicated in the legends of these Figures
and Table 1. A resonance appears for $\omega _{m}\simeq \omega _{ax}$. To
understand the properties of this resonance, one can derive a simplified
analytic expression of $I_{1}$, for a range of frequencies $\left\vert
\omega _{m}-\omega _{ax}\right\vert \ll \Lambda _{a}$, $\Lambda _{m}$:%
\begin{align}
I_{1} &  \simeq 16\Omega_{ax}\frac{K_{am}\varepsilon
_{a}\varepsilon_{m}}{\Lambda_{a}^{2}\Lambda_{m}^{2}}\frac{1}{(\omega
_{m}-\omega_{ax})^{2}+\varkappa_{ax}^{2}}\frac{1}{(1+3J_{a}^{2})\sqrt
{1+J_{a}^{2}}}\label{168}\\
&  \times(\cos(\varphi_{2}-\varphi_{ax}^{0})\left\{  \varkappa_{ax}%
+e^{-\varkappa_{ax}T}((\omega_{m}-\omega_{ax})\sin[T(\omega_{m}-\omega
_{ax})]-\varkappa_{ax}\cos[T(\omega_{m}-\omega_{ax})]\right\}  \nonumber\\
&  +\sin(\varphi_{2}-\varphi_{ax}^{0})\left\{  (\omega_{m}-\omega
_{ax})(1-e^{-\varkappa_{ax}T}\cos[T(\omega_{m}-\omega_{ax})])+e^{-\varkappa
_{ax}T}\varkappa_{ax}\sin[T(\omega_{ax}-\omega_{m}))\right\}  )\nonumber
\end{align}

In the above expression, the phase $\varphi _{ax}^{0}$ is the
value of $\varphi _{ax}$ at the time $t=0$ where the integration of the
signal starts. From Eq.(\ref{168}), the difference between $\varphi
_{ax}^{0} $ and the phase $\varphi _{2}$ of the drive applied to the
gyromagnetic mode is a crucial parameter. For the moment, we will assume
that this parameter can be controlled experimentally and that the signal $%
I_{1}$ can be measured versus $\omega _{m}$ using the same value of $\varphi
_{2}-\varphi _{ax}^{0}$ for all data points. This is not a priori trivial
because the initial integration time for which $\varphi _{ax}^{0}$ has to be
defined changes for each data point. We will discuss at the end of this
section simple ways to solve this problem.

When $\varphi _{2}=\varphi _{ax}^{0}$, the third line of Eq.(\ref{168})
cancels, and the curve $I_{1}$ versus $\omega _{m}$ shows a symmetric peak
with a maximum for $\omega _{m}=\omega _{ax}$ (see top right panel in Fig.2 and
all panels in Fig.3). This is not true anymore when $\varphi _{2}\neq
\varphi _{ax}^{0}$, due to the term in $\sin (\varphi _{2}-\varphi_{ax}^{0}) $ in Eq.(\ref{168}), which is odd in $\omega_{m}-\omega _{ax}$
(see left and middle top panels in Fig.2). Hence, below, we assume that we tune the phase $\varphi _{2}$ equal to $\varphi _{ax}^{0}$ to get a more direct access to the system parameters. Importantly, the width of the signal resonance depends on how the integration time $T$ compares with $1/\varkappa_{ax}$. In the limit where $T\ll 1/\varkappa _{ax}$, the signal $I_{1}$
shows a peak with a width given by $\sim T^{-1}$ over a background of
oscillations due to the $\mathrm{sinc}[T(\omega _{m}-\omega _{ax})]$
component emerges in $I_{1}$ for $\Delta \omega _{m}\gg \varkappa _{ax}$
(see top left panel in Fig. 3). In the opposite limit $1/\varkappa _{ax}\ll T$,
Eq.(\ref{168}) gives
\begin{equation}
I_{1}\simeq 16F(\varepsilon _{a})\frac{\Omega _{ax}K_{am}\varepsilon
_{a}\varepsilon _{m}}{\Lambda _{a}^{2}\Lambda _{m}^{2}}\frac{\varkappa _{ax}}{(\omega _{m}-\omega _{ax})^{2}+\varkappa _{ax}^{2}}  \label{169}
\end{equation}
with
\begin{equation}
F(\varepsilon _{a})=\frac{1}{(1+3J_{a}^{2})\sqrt{1+J_{a}^{2}}}
\end{equation}%
From Eq.(\ref{169}), in the limit of a long integration time, $I_{1}$ shows
a peak at $\omega _{m}=\omega _{ax}$ with a width set by $\varkappa _{ax}$
(see right panel in Fig. 3). This regime gives a direct access to the axions
linewidth. Note that due to the assumption $\left\vert \omega _{m}-\omega
_{ax}\right\vert \ll \Lambda _{a}$, $\Lambda _{m}$ used to derive Eqs.(\ref%
{168}) and (\ref{169}), the above discussion is valid when $\varkappa
_{ax},1/T\ll \Lambda _{a},\Lambda _{m}$, which is realistic for the values
expected for $\varkappa _{ax}$, and compatible with previous hypotheses. For
completeness, we also give the expression of $I_{1}$ in resonant conditions $%
\omega _{ax}=\omega _{m}=\omega _{2}$, $\omega _{1}=\tilde{\omega}_{a}$ and $%
\varphi _{2}=\varphi _{ax}^{0}$, for any $T$:%
\begin{equation}
I_{1}=\frac{16}{(1+3J_{a}^{2})\sqrt{1+J_{a}^{2}}}\frac{1-e^{-\varkappa
_{ax}T}}{\varkappa _{ax}}\frac{\Omega _{ax}K_{am}\varepsilon _{a}\varepsilon
_{m}}{\Lambda _{a}^{2}\Lambda _{m}^{2}}  \label{IresBis}
\end{equation}%
This expression shows that to get an access to $\varkappa _{ax}$, an
alternative protocol could be to measure the dependence of $I_{1}$ with $T$.

\begin{figure}[h]
\begin{center}
\includegraphics[width=0.9\linewidth,angle=0]
{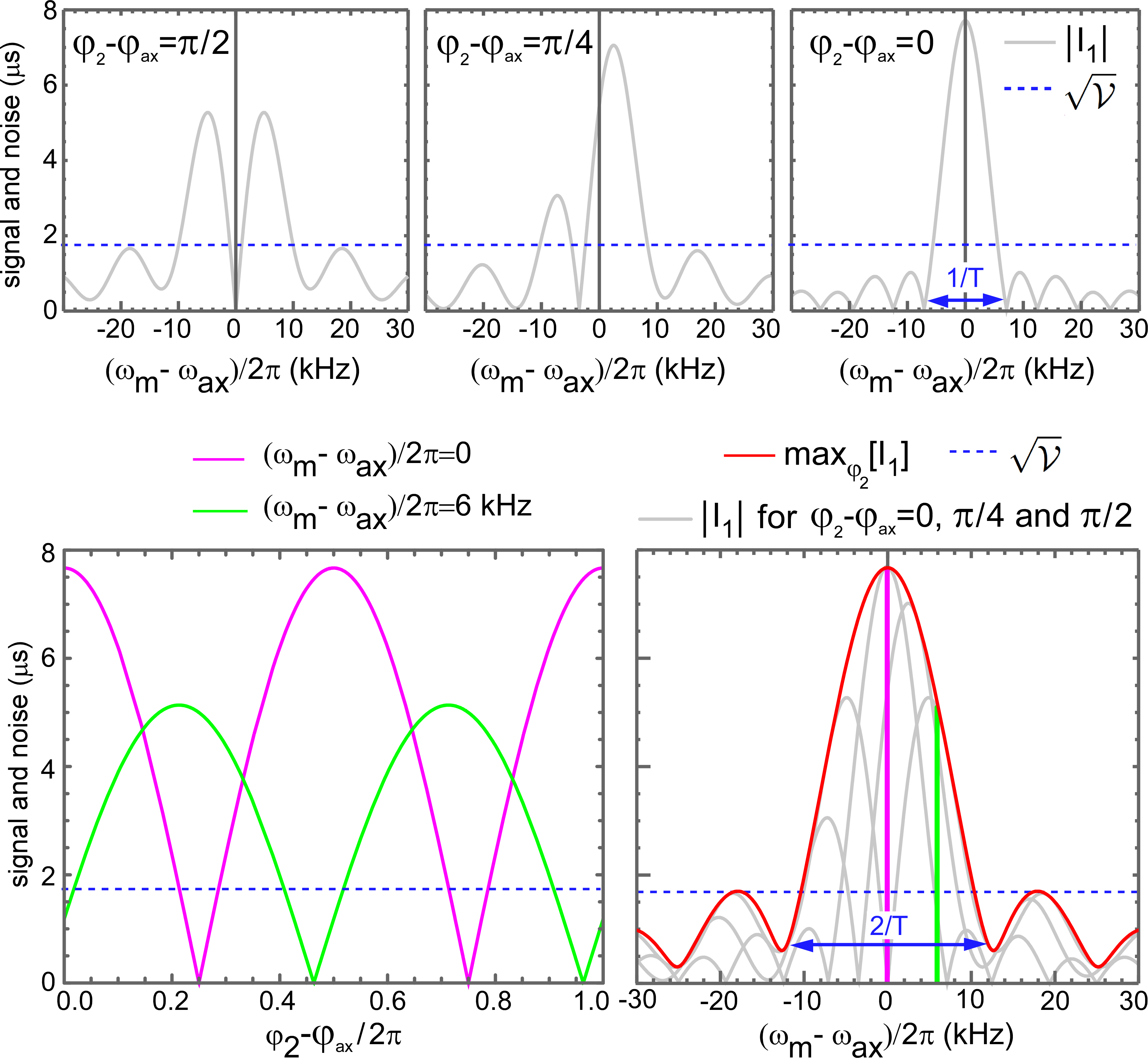}\caption{\textbf{Tuning the phase of the phase resolved haloscope}. Top panels: Axion
signal $\left\vert I_{1}\right\vert $ versus $\omega_{m}-\omega_{ax}$\ for $T=1/2\varkappa_{ax}$, $\varepsilon_{a}=\varepsilon_{a}^{\max}$, different values of $\varphi_{2}$ and the parameters of Table 1 (grey full lines).Bottom left panel: Variations of $\left\vert I_{1}\right\vert $ with $\varphi_{2}-\varphi_{ax}$ for different values of $\omega_{m}-\omega_{ax}$ (pink and green full lines). Bottom right panel: Amplitude of variation of $\left\vert I_{1}\right\vert $\ with  $\omega_{m}-\omega_{ax}$ (red full line). In all the panels, the square root of the detector noise$\sqrt{\mathcal{V}}$ is indicated with blue dashed lines. In the bottom right panel, the curves of the other panels are recalled with a consistent color code.}
\label{Figure2}
\end{center}
\end{figure}

The above results allow us to envision a practical protocol aimed at
detecting possible axions. The first step is to use $\omega _{1}=\omega _{a}$
and $\omega _{2}=\omega _{m}$, and to measure $I_{1}$ versus $\omega _{m}$
with a short integration time $T\ll 1/\varkappa _{ax}$ in order to make a
fast scan of the frequency space available for $\omega _{m}$. If axions do
exist, a resonance should appear around $\omega _{m}\simeq \omega _{ax}$,
with a peak which is not necessarily symmetric because $\varphi _{2}-\varphi
_{ax}^{0}$ will be unknown during the first frequency scan and it can vary
along the measurement if a very wide $\omega _{m}$ range is explored. In a
second step, one should focus on the range of $\omega _{m}\in \lbrack \omega
_{ax}-T^{-1},\omega _{ax}+T^{-1}]$ yielding the axion resonance, using again
a short integration time and changing $\varphi _{2}$ until a symmetric
resonance is obtained, which confirms $\varphi _{2}=\varphi _{ax}^{0}$.
Assuming we take $N$ data points along the resonance, one should use $T\ll 1/N\varkappa _{ax}$. Once the limit $\varphi _{2}=\varphi
_{ax}^{0} $ is achieved with a short time $T$, one can launch without
waiting a measurement for one data point of the curve $I_{1}(\omega _{m})$
with $\varphi _{2}=\varphi _{ax}^{0}$ and $T\gg 1/\varkappa _{ax}$. The
phase tuning of the system for short time $T$ should be repeated between
each measurement of a point with $T\gg 1/\varkappa _{ax}$ in order to ensure
$\varphi _{2}=\varphi _{ax}^{0}$ for each point. This way, the full
resonance in the $I_{1}(\omega _{m})$ curve for $\varphi _{2}=\varphi
_{ax}^{0}$ and $T\gg 1/\varkappa _{ax}$ should be accessible, with a width
revealing the exact value of $\varkappa _{ax}$. Alternatively, the value of $\varkappa _{ax}$ can be obtained by measuring the dependence of $I_{1}$
versus $T$ for $\omega _{ax}=\omega _{m}=\omega _{2}$, $\omega _{1}=\tilde{\omega}_{a}$ and $\varphi _{2}=\varphi _{ax}^{0}$. At last, one could obtain
the amplitude $\Omega _{ax}$ of the axion coupling term from the height of
the $I_{1}$ resonance if the other system parameters $K_{a}$, $K_{am}$, $%
\varepsilon _{a} $, $\varepsilon _{m}$, $\Lambda _{a}$, $\Lambda _{m}$ and $%
\varkappa _{ax}$ are known

In practice, sweeping the frequency $\omega_{m}$ rapidly to measure the $I_{1}(\omega_{m})$ curve might turn out to be technically challenging. In order to circumvent this problem, one may alternatively work for a fixed frequency $\omega_{2}=\omega_{m}$ and sweep rapidly the phase $\varphi_{2}$ of the parametric drive, by using for example analog phase shifters, in order to detect the oscillations of $I_{1}$ with $\varphi_{2}$ (see Figure 2, bottom left panel). This method would require an experimental setup with enough phase stability and would need a careful handling of the drive phase in order to avoid potential added noise by the extra control knob added to the setup. Using a short integration time time $T\approx1/N\kappa_{ax}$, one should be able to get N data points along a $I_{1}(\varphi_{2})$ curve for a fixed $\omega_{2}=\omega_{m}$ . One could then extract for each frequency $\omega_{2}=\omega_{m}$ the maximum value
\begin{equation}
\max\nolimits_{\varphi_{2}}[I_{1}]\simeq16\Omega_{ax}\frac{K_{am}%
\varepsilon_{a}\varepsilon_{m}}{\Lambda_{a}^{2}\Lambda_{m}^{2}(1+3J_{a}^{2})}\sqrt{\frac{1+e^{-2\varkappa_{ax}T}+e^{-\varkappa_{ax}T}\cos[T(\omega_{m}-\omega_{ax})]}{(1+J_{a}^{2})((\omega_{m}-\omega_{ax})^{2}+\varkappa_{ax}^{2})}}
\end{equation}
reached by $I_{1}(\varphi_{2})$, in order to reconstruct the red curve shown in Figure 2, bottom right panel. For a quick integration time $T\ll1/\kappa_{ax}$, this curve shows a peak at $\omega_{2}=\omega_{ax}$ with a width $1/2T$. This can be used to determine the frequency $\omega_{2}=\omega_{ax}$. From this point, the detection of the value of $\varkappa_{ax}$ can be obtained for instance by measuring the dependence of $I_{1}$ versus $T$, which follows Eq.(\ref{IresBis}). As a final note of this section, it is worth mentioning that a short integration time can be used here as a resource to compensate the very narrow bare linewidth expected for axion signals since their apparent linewidth becomes $T^{-1}$ in this limit. This can be particularly useful when a wide frequency range is used to span the large coupling/mass axion plane.

\begin{figure}[h]
\begin{center}
\includegraphics[width=1.0\linewidth,angle=0]{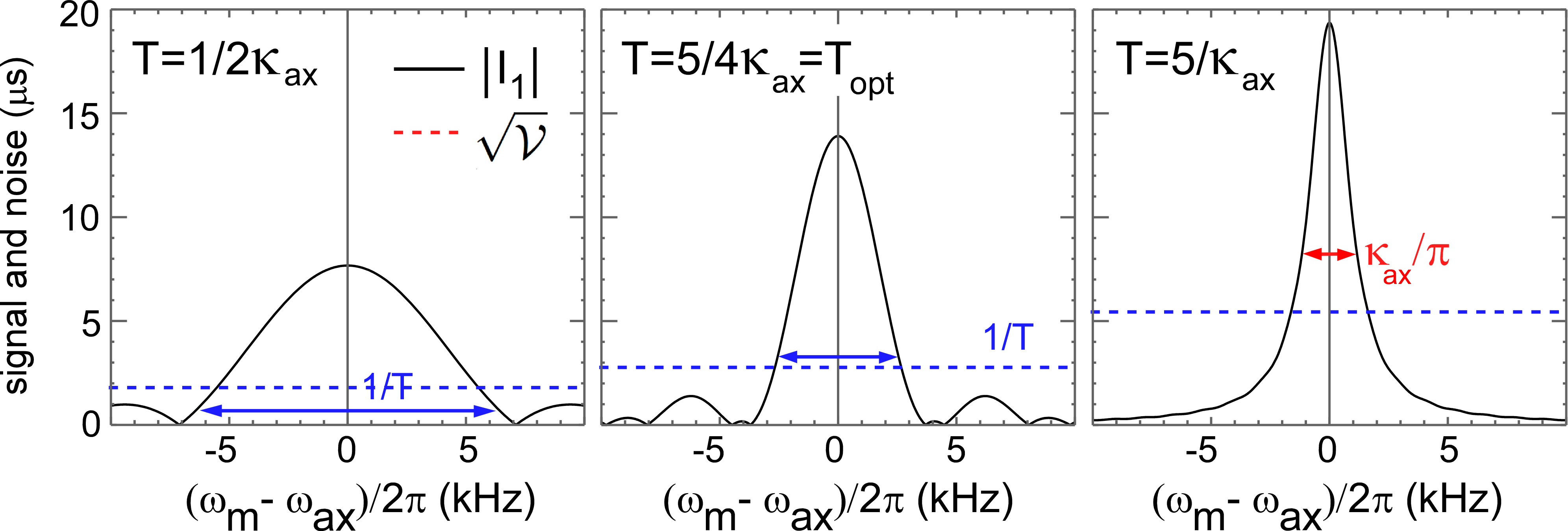}
\caption{\textbf{Choosing the integration time of the phase resolved haloscope}. Axion signal $I_{1}$\ and square root $\sqrt{\mathcal{V}}$ of the detector noise versus $\omega _{m}-\omega _{ax}$\ for different values of $T$, $\varepsilon _{a}=\varepsilon _{a}^{\max }$, $\varphi _{2}=\varphi_{ax}^{0}$ and the parameters of Table 1. The timing of the measurement can be used here as a resource to compensate the very narrow linewidth expected for axion signals by shortening the measurement time compared to the axion decoherence time.}
\label{Figure3}
\end{center}
\end{figure}

\textbf{Signal-to-noise ratio of the phase-resolved haloscope at
resonance\label{SignalRes}}

To benchmark the phase-resolved haloscope, one must calculate the signal
variance, which is defined by Eq.(\ref{Vexpr}). The comparison between the
signal and its variance will set the range of integration times $T$ which
can be used. The crucial question to answer is whether the limits of
integration times $NT\ll 1/\varkappa _{ax}$ and $T\gg 1/\varkappa _{ax}$
discussed in the previous section will give an effective access to $I_{1}$
thanks to a signal-to-noise ratio greater than unity.

To understand the noise properties of the phase-resolved haloscope, one
needs to calculate the two-time correlations of the cavity field. For this
purpose, one can quadratize the effective cavity action with respect to the
photonic fields, assuming that the photonic behavior of the cavity is not
strongly nonlinear. This requires to use a drive amplitude $\varepsilon _{a}$
smaller than the critical drive $\varepsilon _{a}^{crit}$ above which the
readout mode shows a hysteretic behavior when $\omega _{1}$ is swept (see SI Eq. (125)). Then, one can map the resulting action onto a Lindblad
description which is suitable for studying the time-dependent behavior of
the detector. Finally, we calculate the correlation functions of the field
in the readout mode by using a linear regression theorem which is valid
because we treat the dissipation in the system with a Markovian
approximation. These steps are presented in details in SI, and the general expression of $\mathcal{V}$ for $\varphi _{d}=\varphi _{d}^{\pm }$
is given in Eq.(218) of the SI. In resonant conditions $\omega _{1}=\tilde{\omega}_{a}$ corresponding to our protocol, one obtains:%
\begin{equation}
\mathcal{V}=\frac{(1+2n_{B}(\omega _{a}))T}{\Lambda _{a}}\frac{(1+9J_{a}^{2})%
}{(1+3J_{a}^{2})^{2}}  \label{Vres}
\end{equation}%
Note that due to the quantumness of our phase-resolved haloscope, $\mathcal{V%
}$ differs from the classical prefactor $\mathcal{V}_{class}=(1+2n_{B}(\omega _{a}))T/\Lambda _{a}$ since the function $%
(1+9J_{a}^{2})/(1+3J_{a}^{2})^{2}$ can be larger or smaller than 1 (see Fig.5b). The value of $\sqrt{\mathcal{V}}$ for $\omega _{1}=\tilde{\omega}%
_{a}$ is shown in each panel of Figs. 2 and 3 with blue dashed lines. A resolvable signal $I_{1}>%
\sqrt{\mathcal{V}}$ is obtained when $\omega _{m}\simeq \omega _{ax}$. Using
Eqs. (\ref{IresBis}) and (\ref{Vres}), the signal-to-noise ratio
\begin{equation}
\mathcal{R}=I_{1}/\sqrt{\mathcal{V}}
\end{equation}%
can be expressed as%
\begin{equation}
\mathcal{R}=\frac{\Omega _{ax}K_{am}\varepsilon _{a}\varepsilon _{m}}{\sqrt{\varkappa _{ax}}\Lambda _{a}^{3/2}\Lambda _{m}^{2}}\frac{16R_{ax}(T)}{\sqrt{(1+2n_{B}(\omega _{a}))}}\frac{1}{\sqrt{(1+J_{a}^{2})(1+9J_{a}^{2})}}
\label{RRR}
\end{equation}%
with%
\begin{equation}
R_{ax}(T)=\frac{1-e^{-\varkappa _{ax}T}}{\sqrt{\varkappa _{ax}T}}
\end{equation}%
Importantly, equation (\ref{RRR}) shows that one can optimize $\mathcal{R}$
by tuning both the amplitudes $\varepsilon _{a}$ and $\varepsilon _{m}$ of
the microwave excitations applied to the cavity and gyromagnetic modes. To
get a large signal, it is advantageous to increase $\varepsilon _{m}$ as
much as possible, but one should be careful to remain in the regime $\Lambda
_{\phi }\ll \Lambda _{a}$ to avoid measurement-induced dephasing. Besides,
the dependence of the signal-to-noise ratio on $\varepsilon _{a}$ is not
trivial due to the presence of the $J_{a}$ factors in the expression of $%
I_{1}$. We will address these matters in the sections that follow.

\textbf{Integration time $T_{opt}$ to maximize the signal-to-noise ratio}

From Eqs.(\ref{IresBis}), (\ref{Vres}) and (\ref{RRR}), the functional
dependences of $I_{1}$, $\mathcal{V}$ and $\mathcal{R}$ on the integration
time $T$ are independent from the values of $\varepsilon _{a}$ and $%
\varepsilon _{m}$. The time $T$ which optimizes the signal-to-noise ratio of
the phase-resolved haloscope just depends on the value of the axion
linewidth $\varkappa _{ax}$. More precisely, in Eq.(\ref{RRR}), the function
$R_{ax}(T)$ which sets the variations of $\mathcal{R}$ with $T$ reaches a
maximum%
\begin{equation}
R_{ax}(T_{\text{opt}})=\sqrt{-\frac{1}{2}\mathcal{L}(-1,-\frac{1}{2\sqrt{e}}%
)^{-2}-\mathcal{L}(-1,-\frac{1}{2\sqrt{e}})^{-1}}
\end{equation}%
for the integration time $T=T_{\text{opt}}$ with
\begin{equation}
T_{\text{opt}}=-\frac{1}{2\varkappa _{ax}}(1+2\mathcal{L}(-1,-\frac{1}{2%
\sqrt{e}})\simeq \frac{5}{4\varkappa _{ax}}  \label{Topt}
\end{equation}%
Above, $\mathcal{L}$ is the Lambert function also called product logarithm.
In Fig.4, the maximum of $\mathcal{R}$ at $T=T_{\text{opt}}$ is marked by
the vertical magenta line. One can check numerically that $R_{ax}(T_{\text{opt}})\simeq 0.64$. For the parameters of Fig. 4 and $T=T_{\text{opt}}=0.2~%
\mathrm{ms}$, one obtains $I_{1}=14~\mathrm{\mu s}$ and $\mathcal{R}=5.2$.

\begin{figure}[h]
\begin{center}
\includegraphics[width=0.8\linewidth,angle=0]
{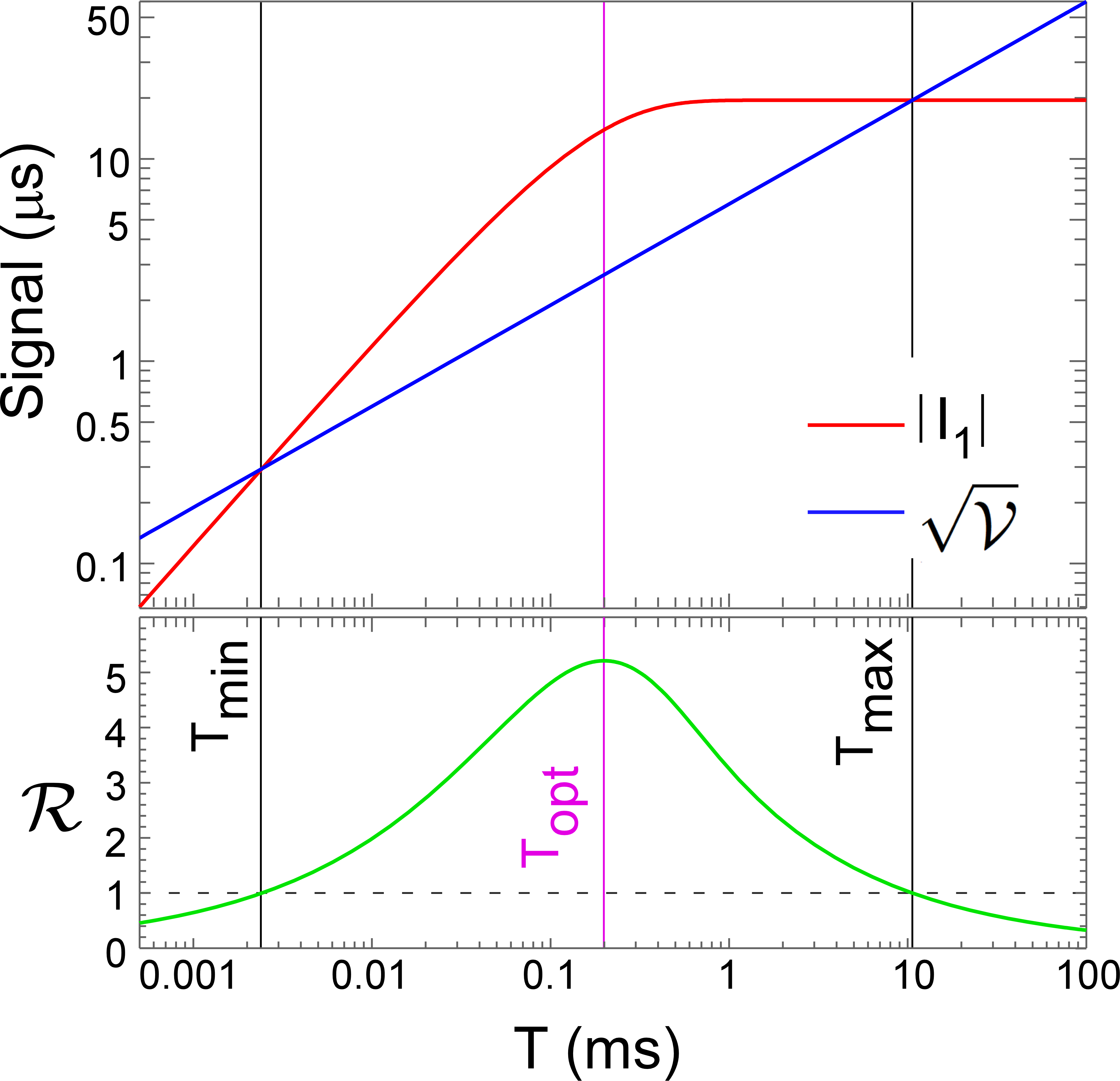}\caption{\textbf{Optimum measurement time.}. Axion signal $I_{1}$\ (top panel, red curve) square root of detector variance $\sqrt{\mathcal{V}}$\ (top panel, blue curve), and signal to noise ratio $R$\ (bottom panel, green curve) versus the integration time $T$\ for $\varepsilon _{a}=\varepsilon _{a}^{\max }$, $\omega _{1}=\tilde{\omega}_{a}$, $\omega _{2}=\omega _{m}=\omega _{ax}$, $\varphi _{2}=\varphi_{ax}^{0}$, and the parameters of Table 1, including $\varepsilon _{m}\simeq\varepsilon _{m}^{\max }$. For the specific parameters chosen here, we get an optimum measurement time $T_{\text{opt}}=200 \mu s$.}
\label{Figure4}
\end{center}
\end{figure}

\textbf{Time window $[T_{\min },T_{\max }]$ to obtain a signal-to-noise ratio larger than unity}

The range of possible integration times giving $\mathcal{R}\geq 1$ is a
crucial feature to characterize the figure of merit of the phase-resolved
haloscope. For times $T\ll 1/\varkappa _{ax}$, the signal $I_{1}$ is a
linear function of $T$ whereas for $T\gg 1/\varkappa _{ax}$, $I_{1}$ tends
to a constant value. We assume that the $I_{1}$ curve intersects $\sqrt{%
\mathcal{V}}$ in these two regimes and we note $T_{\min }$ and $T_{\max }$
the corresponding times. This situation is illustrated by Fig. 4, top panel.
Using $\omega _{ax}=\omega _{m}=\omega _{2}$, $\omega _{1}=\tilde{\omega}%
_{a} $ and $\varphi _{2}=\varphi _{ax}^{0}$, one finds:

\begin{equation}
T_{\min }=(1+2n_{B})\frac{(1+J_{a}^{2})(1+9J_{a}^{2})}{256}\frac{\Lambda
_{a}^{3}\Lambda _{m}^{4}}{\Omega _{ax}^{2}K_{am}^{2}\varepsilon
_{a}^{2}\varepsilon _{m}^{2}}  \label{Tmin}
\end{equation}%
and%
\begin{equation}
T_{\max }=\frac{1}{\varkappa _{ax}^{2}T_{\min }}  \label{TmaxBis}
\end{equation}
The value of $T_{\text{opt}}$ must necessarily lie between $T_{\min }$ and $T_{\max }$ when the definition of these quantities is relevant. One can
indeed check from Eqs.(\ref{RRR}), (\ref{Topt}), (\ref{Tmin}), and (\ref{TmaxBis}) that if $\mathcal{R}>1$ for $T=T_{\text{opt}}$, the inequalities $T_{\min }<T_{\text{opt}}<T_{\max }$ are satisfied (see details in SI, section V.6). Nevertheless the above expressions of $T_{\min }$ and $T_{\max }$
correspond to the true integration times $\tilde{T}_{\min }$ and $\tilde{T}%
_{\max }$ which yield $\mathcal{R}=1$ only if the intersections of the $%
I_{1} $ and $\sqrt{\mathcal{V}}$ curves do occur in the linear and constant
regime of $I_{1}$. This happens for the parameters used in Fig. 4, for which
one gets $T_{\min }=2.4~\mathrm{\mu s}$, $T_{\max }=10.6~\mathrm{ms}$ and $%
T_{\text{opt}}=0.2~\mathrm{ms}$. This is also correct for most of the range
of $\varepsilon _{a}$ considered in Fig.5d (compare red full line and red
dots). Equation (\ref{Tmin}) is the main result of our work since it sets
the fidelity $F_{phase}=1/T_{\min }$ of the phase resolved haloscope.

\textbf{Influence of the drive parameter $\protect\varepsilon _{a}$\label{optreg}}
\begin{figure}[h!]
\begin{center}
\includegraphics[width=0.35\linewidth,angle=0]
{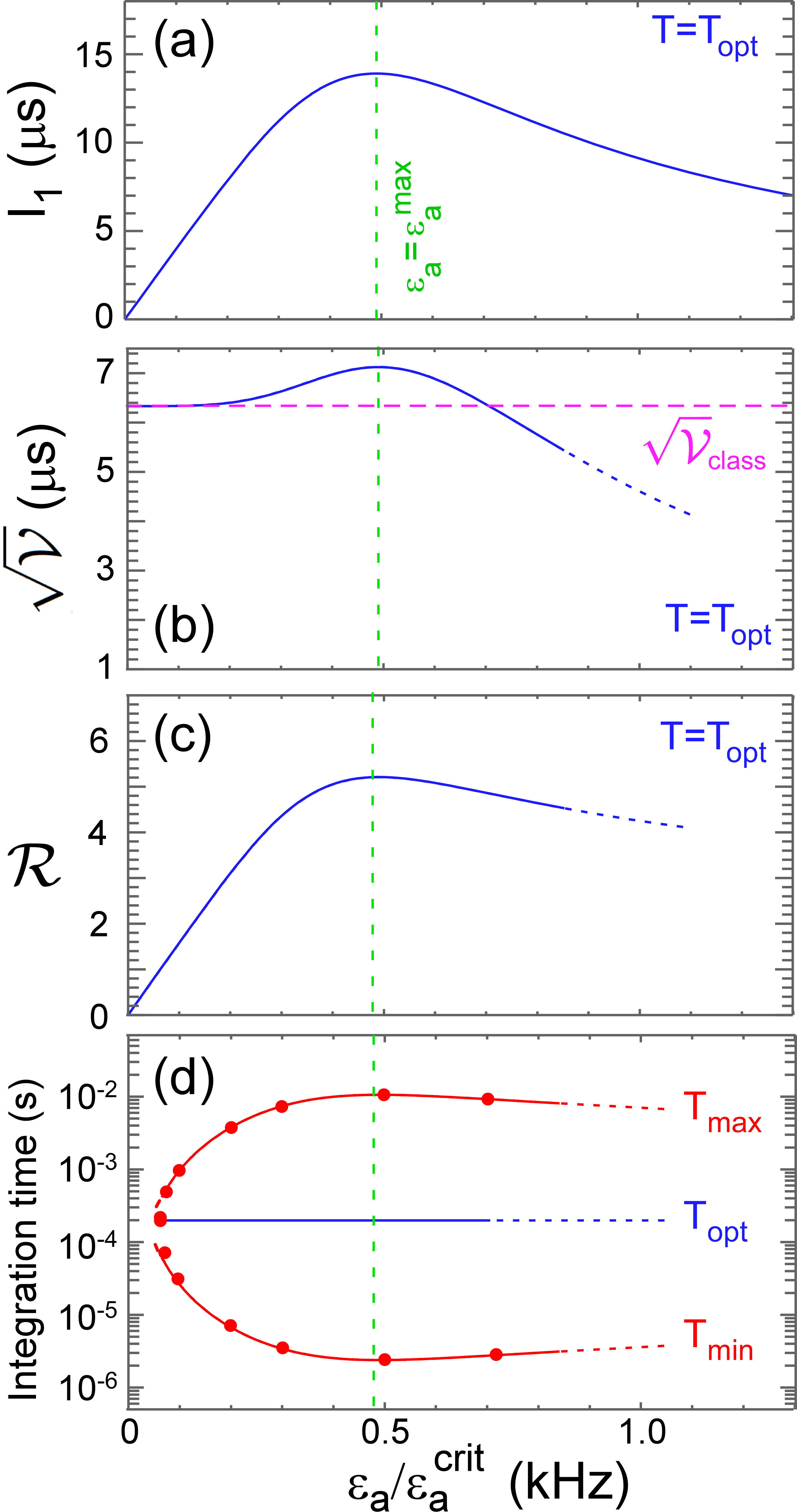}\caption{\textbf{Optimum working point.} Axion signal $I_{1}$(panel \textbf{a}), square root of the detector
variance $\sqrt{\mathcal{V}}$\ (panel \textbf{b}), and signal-to-noise ratio $\mathcal{R}$\
(panel \textbf{c}); integration times $T_{\min }$, $T_{\text{opt}}$\ and $T_{\max }$\
(panel \textbf{d}) as a function of the excitation amplitude $\varepsilon _{a}$\ for $\omega _{1}=\tilde{\omega}_{a}$, $\omega _{2}=\omega _{m}=\omega _{ax}$, $\varphi _{2}=\varphi _{ax}^{0}$\ and the parameters of Table 1, including $%
\varepsilon _{m}\simeq \varepsilon _{m}^{\max }$. In panels \textbf{(a)}, \textbf{(b)} and
\textbf{(c)}, the values of $\left\vert \mathcal{I}_{1}^{\pm }\right\vert $, $\sqrt{\mathcal{V}}$\ \ and $\mathcal{R}$\ are shown for $T=T_{opt}$. We have used $K_{a}=-2\pi \times 0.212$~Hz . The classical value $V_{class}$ of the variance
is indicated as a magenta horizontal line in panel \textbf{(c)}. We show $\sqrt{\mathcal{V}}$ and $\mathcal{R}$ only for moderate values of $\varepsilon _{a}$ for which the mathematical approach presented in the SI is valid. In panel \textbf{(d)}, the full red lines correspond to the times $T_{\min }$\ and $T_{\max }$\ given by the equations (\ref{Tmin}) and (\ref{TmaxBis}) and the dots correspond to
the exact integrations times such that $\mathcal{R}=1$, calculated numerically. The value of $T_{\text{opt}}$\ is independent from the value of $K_{a}$.}
\end{center}
\end{figure}
We now search for the optimal value of $\varepsilon _{a}$. We assume again $\omega _{ax}=\omega _{m}=\omega _{2}$, $\omega _{1}=\tilde{\omega}_{a}$ and $\varphi _{2}=\varphi _{ax}^{0}$. Figure 5 shows the variations of $I_{1}$, $\mathcal{V}$, and $\mathcal{R}$ with $\varepsilon _{a}$ for $T=T_{\text{opt}} $ It also shows the variations of $T_{\min }$ and $T_{\max }$.
Interestingly, for large values of $\varepsilon _{a}$, the variance $%
\mathcal{V}$ becomes smaller than its classical value due to quantum
squeezing. From Eqs. (\ref{IresBis}), (\ref{Vres}) and (\ref{RRR}), one can
check analytically that $I_{1}$, $\mathcal{V}$, $\mathcal{R}$, $T_{\min
}^{-1}$ and $T_{\max }$ are maximal for $\varepsilon _{a}=\varepsilon
_{a}^{\max }$ with%
\begin{equation}
\varepsilon _{a}^{\max }=\frac{1}{6^{3/2}}\sqrt{\frac{5\Lambda _{a}^{3}}{%
\left\vert K_{a}\right\vert }}\simeq 0.5\varepsilon _{a}^{crit}  \label{optP}
\end{equation}%
with $\varepsilon _{a}^{crit}$ the critical drive amplitude of the Josephson
junction, defined in Eq.(125). The optimal value $\varepsilon
_{a}^{\max }$ thus corresponds to a non-hysteretic cavity behavior. For $%
\varepsilon _{a}=\varepsilon _{a}^{\max }$, one gets, for any value of $T$%
\begin{equation}
I_{1}=\sqrt{3}\frac{1-e^{-\varkappa _{ax}T}}{\varkappa _{ax}}\frac{\Omega
_{ax}K_{am}\varepsilon _{m}}{\sqrt{\Lambda _{a}\left\vert K_{a}\right\vert }%
\Lambda _{m}^{2}}  \label{Iopt}
\end{equation}%
\begin{equation}
\mathcal{V}=\frac{9}{8}\frac{(1+2n_{B}(\omega _{a}))T}{\Lambda _{a}}
\end{equation}%
and%
\begin{equation}
\mathcal{R}=2\sqrt{\frac{2}{3}}R_{ax}(T)\frac{\Omega _{ax}K_{am}\varepsilon
_{m}}{\Lambda _{m}^{2}}\frac{1}{\sqrt{\varkappa _{ax}\left\vert
K_{a}\right\vert (1+2n_{B}(\omega _{a}))}}  \label{Ropt}
\end{equation}%
The value of $T_{\min }$ obtained for $\varepsilon _{a}=\varepsilon
_{a}^{\max }$, is
\begin{equation}
T_{\min }=\frac{3(1+2n_{B}(\omega _{a}))\left\vert K_{a}\right\vert \Lambda
_{m}^{4}}{8K_{am}^{2}\varepsilon _{m}^{2}\Omega _{ax}^{2}}
\end{equation}%
and the corresponding value for $T_{\max }$ derives from Eq.(\ref{TmaxBis}).
Paradoxically, the value of $\varepsilon _{a}$ which optimizes $\mathcal{R}$
corresponds to a value of $\mathcal{V}$ slightly larger than the classical
value $\mathcal{V}_{class}$, because this is compensated by a large $I_{1}$.
In the above expressions, we have mentioned explicitly the readout mode nonlinearity $K_{a}$. This presentation puts forward an interesting property of our setup: the light/axion coupling occurs through the ratio $\Omega _{ax}^{2}/K_{a}$. Interestingly, the critical drive $\varepsilon _{a}^{\max}$ of the cavity also scales with $K_{a}^{-1}$. This means that if the coupling $\Omega _{ax}$ is too small, this can be compensated to a certain extent by using a smaller value for $K_{a}$ and a larger value for the cavity drive $\varepsilon _{a}$ (as one can anticipate from figure 1b). This shows the robustness of our proposal which is relatively flexible regarding the choice of experimental parameters. This flexibility could be advantageous to push the detection window to lower frequencies, by achieving devices with $K_a$ to compensate for the small couplings $\Omega_{ax}$ at low magnetic fields. Note also that the equation (\ref{Iopt}) can give the impression that the value
of the signal $I_{1}$ diverges at the working point $\varepsilon_{a}=\varepsilon _{a}^{\max }$ for $K_{a}\rightarrow 0$. However, one should keep in mind that from Eq. (\ref{optP}), $\varepsilon _{a}^{\max }$ should also diverge when $K_{a}\rightarrow 0$. This means that when $K_{a}$ is very small, it is necessary to apply a very strong drive $\varepsilon _{a}$ to reach the $\varepsilon _{a}=\varepsilon _{a}^{\max }$ working point. In practice, higher-order non linear effects and heating effects should regularize this regime which is beyond the scope of the our study. Nevertheless, for a cavity coupled to a non-linear Josephson junction, a finite value of $K_{a}$ is expected. Note that for $\varepsilon _{a}=\varepsilon_{a}^{\max }$ and the resonant conditions described above, the average number of photons in the cavity is about $2\times 10^6$. In this limit, the semiclassical approximation $\left\langle \hat{a}(t)\right\rangle =a(t)$
that we have used to calculate $I_{1}$ is definitely valid.

\textbf{Influence of the drive parameter $\protect\varepsilon _{m}$\label%
{choiceEpsm}}

From Eq.(\ref{Tmin}), it seems favorable to use a large value of $%
\varepsilon _{m}$ to optimize the figure of merit of the phase-resolved
haloscope. In this limit, one has to be careful about the contributions
proportionnal to $K_{am}^{2}\varepsilon _{m}^{2}$ in the readout mode
effective action. In the regime $\omega _{2}=\omega _{m}$ used in our
protocol, the contribution in $K_{am}^{2}\varepsilon _{m}^{2}$ to $%
\widetilde{K}_{a}$ cancels (see Eq.\ref{Katild}). However, the dephasing
rate $\Lambda _{\phi }$ of Eq.(\ref{LandaPhi}) is finite. A simple criterion
for the proper operation of the phase-resolved haloscope is $\Lambda _{\phi
}\ll \Lambda _{a}$. Here, we will impose $\Lambda _{\phi }=\Lambda _{a}/10$.
This implies a drive value $\varepsilon _{m}=\varepsilon _{m}^{\lim }$ such
that%
\begin{equation}
\varepsilon _{m}^{\lim }=\frac{1}{4}\sqrt{\frac{\Lambda _{a}\Lambda _{m}^{3}%
}{10K_{am}^{2}(1+2n_{B}(\omega _{m}))}}
\end{equation}%
Using the drive amplitudes $\varepsilon _{a}=\varepsilon _{a}^{\max }$ and $%
\varepsilon _{m}=\varepsilon _{m}^{\lim }$, and assuming $T=0$ (so that $%
n_{B}(\omega _{a/m})=0$) for simplicity of the expressions, one finally
gets:
\begin{equation}
\mathcal{R}(T)=R_{ax}(T)\frac{\Omega _{ax}}{\sqrt{\varkappa _{ax}\left\vert
K_{a}\right\vert }}\sqrt{\frac{\Lambda _{a}}{\Lambda _{m}}}\sqrt{\frac{1}{15}%
}
\end{equation}%
and
\begin{equation}
T_{\min }=\frac{15\left\vert K_{a}\right\vert \Lambda _{m}}{\Omega
_{ax}^{2}\Lambda _{a}}
\end{equation}%
These equations suggest that it is useful to use $\Lambda _{m}$ smaller that
$\Lambda _{a}$ to optimize the performances of the detector. Note that the
choice of parameters in Table 1 is a bit more conservative than $\Lambda
_{\phi }<\Lambda _{a}/10$ since we use $\Lambda _{\phi }=0.4MHz$ and $%
\Lambda _{a}=5MHz$.

\textbf{Quantitative predictions for the figure of merit of the phase-resolved haloscope \label{outputSignal}}
In practice, the signal measured experimentally will be the average voltage
at the output of the microwave transmission line connected to the readout
mode, defined by Eq.(\ref{I_out}), while the output signal-to-noise ratio $%
\left\vert \mathcal{I}_{out}\right\vert /\sqrt{\mathcal{V}_{out}}$ is equal
to $\mathcal{R}$. We assume $Z=50~\Omega $, $\tilde{\omega}_{a}=2\pi \times 6
$\textit{~}\textrm{GHz}, \textrm{\ }$K_{a}=-2\pi \times 0.212$\textit{~}\textrm{%
Hz}, $\Lambda _{a}=2\pi \times 5$\textit{~}\textrm{MHz}, $\omega
_{ax}=\omega _{m}=\omega _{2}$, $\omega _{1}=\tilde{\omega}_{a}$, $\varphi
_{2}=\varphi _{ax}^{0}$, an excitation amplitude $\varepsilon
_{a}=\varepsilon _{a}^{\max }$ defined by Eq.(\ref{optP}) for the readout
mode, and all other parameters given in Table 1, including a value $%
\varepsilon _{m}\simeq \varepsilon _{m}^{\lim }$. In this case, the internal
cavity signal $\mathcal{I}=0.29~\mathrm{\mu s}$ obtained for $T=T_{\min }=2.4
$\textit{~}$\mathrm{\mu s}$\textrm{\ }corresponds to a signal-to-noise ratio
$\mathcal{R}=1$, an average external voltage $\mathcal{I}_{out}=9.6~\mathrm{%
nV}$, a power $\mathcal{I}_{out}^{2}/Z=1.8~10^{-18}~\mathrm{W}$\textrm{, and
a figure of merit} $F_{phase}=1/T_{\min }\approx 0.4~MHz$. Importantly, one
finds $T_{\min }\ll \varkappa _{ax}^{-1}$ with $\varkappa _{ax}^{-1}=1$%
\textit{~}$\mathrm{ms}$ in our chosen set parameters. This means that, in
principle, one can reach the short integration time regime discussed previously, where the width of the haloscope resonance is
limited by $T^{-1}$. The internal cavity signal $\mathcal{I}=14~\mathrm{\mu s%
}$ obtained for $T=T_{\text{opt}}=0.2$\textit{~}$\mathrm{ms}$\textrm{\ }%
corresponds to a signal-to-noise ratio $\mathcal{R}=5.2$, an average
external voltage $\mathcal{I}_{out}=5.5~\mathrm{nV}$ and a power $\mathcal{I}%
_{out}^{2}/Z=6~10^{-19}~\mathrm{W}$. Finally, for $T=T_{\max }=11$\textit{~m}%
$\mathrm{s}$, one gets $\mathcal{I}=19~\mathrm{\mu s}$, $\mathcal{I}_{out}=0.14~%
\mathrm{nV}$, a power $\mathcal{I}_{out}^{2}/Z=4.2~10^{-22}~\mathrm{W}$.
Such output signals are detectable with present-day microwave equipment. In
these conditions, reaching the limit $T\gg 1/\varkappa _{ax}$ where the
width of the haloscope resonance is limited by $\varkappa _{ax}^{-1}$ seems
feasible.



\textbf{Figure of merit}

By definition, the figure of merit of a haloscope is the inverse of the
minimum integration time required to achieve a signal-to-noise ratio of $1$.
In the case of conventional haloscopes, this quantity is usually calculated
by considering the Dicke radiometer equation or alike\cite{axionrate:17}. We
can recalculate it explicitly, starting from Eq.(\ref{EffectiveAction}) with
$K_{a}=0$ and $K_{am}=0$ (see details in the SI, section I). We find that the
figure of merit of a conventional haloscope writes:
\begin{equation}
F_{halo}=\frac{(\Omega _{ax}^{0})^{4}}{\Lambda _{a}^{3}(1+2n_{B}(\omega_{a}))^{2}}
\end{equation}%
This result is in agreement with \cite{axionrate:17}. For comparison, we
give the figure of merit $F_{phase}=T_{\min }^{-1}$ of the phase-resolved
haloscope, given by Eq.(\ref{Tmin}):
\begin{equation}
F_{phase}=\frac{\Omega _{ax}^{2}K_{am}^{2}\varepsilon _{a}^{2}\varepsilon
_{m}^{2}}{\Lambda _{a}^{3}\Lambda _{m}^{4}}\frac{256}{(1+2n_{B}(\omega _{a}))%
}\frac{1}{(1+9J_{a}^{2})(1+J_{a}^{2})}  \label{Fphase}
\end{equation}%
We recall that $\Omega _{ax}^{0}$ and $\Omega _{ax}$ are both proportional
to the axion-photon coupling constant $g_{a\gamma \gamma }$. The above
expressions of $F_{halo}$ and $F_{phase}$ show that our phase-resolved
protocol allows one to "pay" only a power 2 penalty on $g_{a\gamma \gamma }$
instead of the power 4 expected for a power measurement. This fact
represents a major advantage of the phase-resolved protocol. We have assumed in our paper that our setup was equipped with a first-stage amplifier adding the lowest noise corresponding to about 1 photon . In order to avoid adding an extra layer of modelling, this can be accounted for by a higher effective temperature in the $(1+2n_B)$ term in the figure of merit of equation (\ref{Fphase}). For example, $(1+2n_B)\approx 2.2$ for a effective temperature of $T_{eff}\approx250mK$ at $5GHz$. In our case, the favorable $1/(1+2n_B)$ scaling for the phase resolved haloscope as opposed to the $1/(1+2n_B)^2$ scaling for the standard haloscope is an interesting aspect since an increase of noise temperature of our first-stage amplifier would degrade only linearly the figure of merit. We have in mind that the full experimental setup will be equipped with quantum limited, or nearly quantum limited amplifiers like Travelling Wave Parametric Amplifiers (TWPAs), which would make the $(1+2n_B)$ term of the order of $1$.  Note that $F_{halo}${} does
not depend on any drive amplitude, since the conventional haloscope is not
externally driven, and the net outgoing power is expected to arise solely
from possible axions. In contrast, as we have seen in Equation (\ref{Fphase}),
$F_{phase}$ can be boosted by the drives in $\varepsilon _{a}$ and $\varepsilon _{m}$. There is however a limitation to this increase. First, the cavity drive $\epsilon_a$ generates squeezing through the self Kerr term $K_a$ in principle, as shown in figure 5 second panel and in the denominator of equation (43) through the J functions. This nonlinear behavior induced by $K_a$ implies that the optimal value for $\epsilon_a$ is $\epsilon_a^{max}$. Further increase is not useful. Second, the magnon drive $\epsilon_m$ has a measurement induced dephasing-like effect $\Lambda_\phi$ through number fluctuations of the coherent state drive. Hence, its value has to be maintained below a certain limit (see Eq. (39)). This is fully accounted for by our systematic expansion of the effective action of the system (see SI section III.3 for more details on the derivation).


The full connection of our work to the literature on conventional haloscopes
is made in the  SI, section I. In order to be more specific about the expectation
for $F_{phase}$, we have used the typical
parameters of the YIG for the magnetic mode, a moderate volume of $2.7~%
\mathrm{l}$ for the estimation of $\Omega _{ax}$ and a moderate magnetic
field of $2~\mathrm{T}$. We find $F_{phase}=1/T_{\min }\approx 0.4~MHz$\textrm{.} For comparison, $F_{halo}$ for similar cavity volume and magnetic field would be below the $10mHz$ range. The typical values for higher
magnetic field and volumes for practical implementations of haloscopes are
about $100~\mathrm{mHz}$ \cite{Brubaker:17,Braine:20,CAPP:22}.

\begin{figure}[h!]
{\small \centering\includegraphics[width=0.95\linewidth,angle=0]{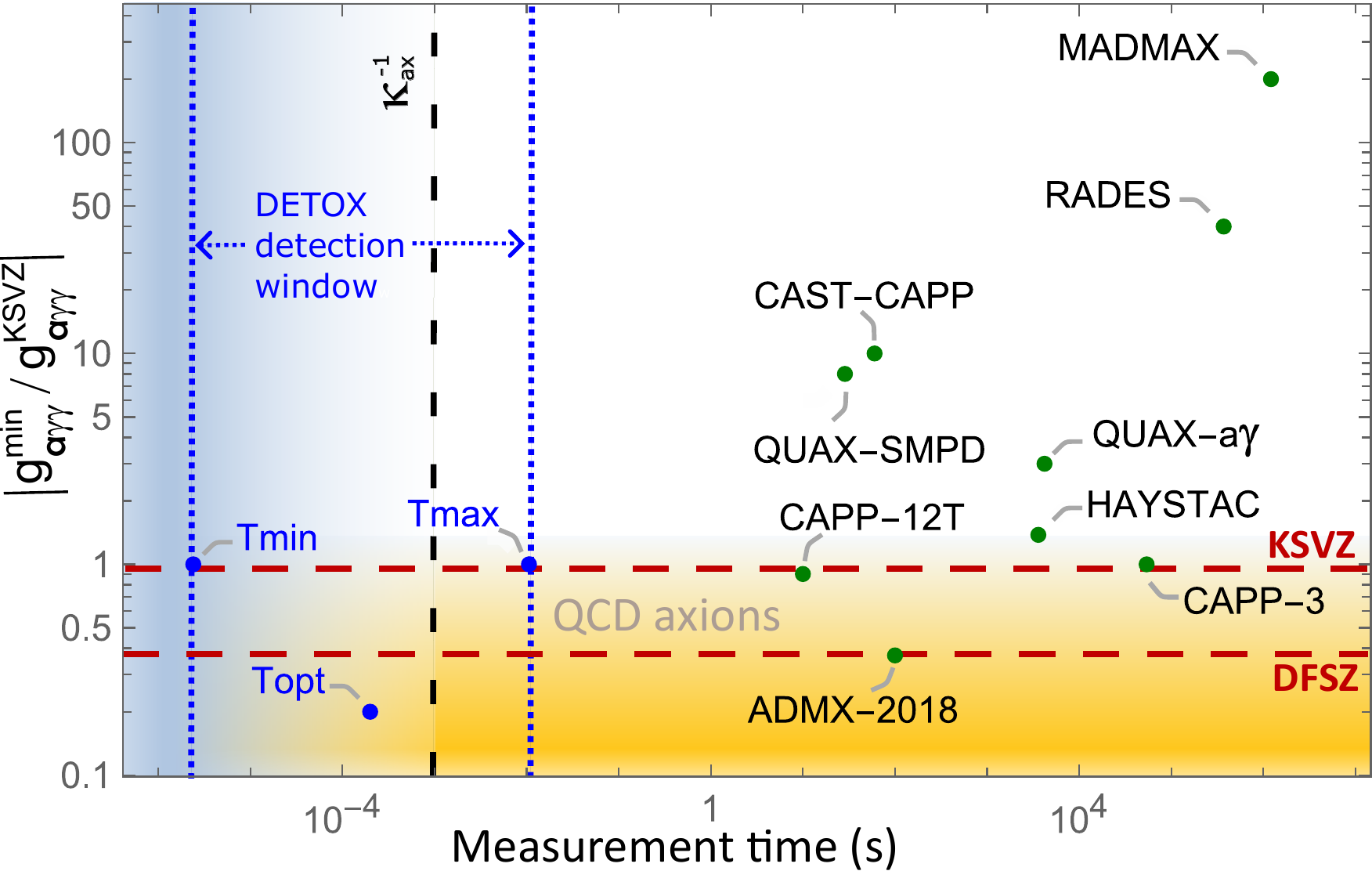}}
\caption{\textbf{State-of-the-Art Haloscope Benchmark}\newline
Summary of the detection times and sensitivity of our phase-resolved
haloscope  proposal named DETOX (blue points) in comparison with some of the main
haloscope experiments of the literature (green points). For our setup, we show the minimum
value $g_{a\gamma \gamma }^{\min }$ detectable for $g_{a\gamma \gamma }$ for
the integration times $T_{\min }$, $T_{\text{opt}}$ and $T_{\max }$ The
values of $g_{a\gamma \gamma }$ expected in the KSVZ and DFSZ models appear as horizontal red dashed lines. The decoherence of $\varkappa _{ax}^{-1}$ of the axion field appears as a vertical black vertical dashed line. While
cosmologically relevant sensitivities have been reached already for narrow mass ranges, the DETOX proposal is the only so far enabling real time
detection.}
\label{fig:figureofmerit}
\end{figure}

It is of interest to benchmark both our projected sensitivity and the
corresponding measurement time against data from existing experiments
reported in the literature\cite{Brubaker:17, Braine:20, Backes:21,
QUAX:24,Caputo:2021}. Each green point in Fig. 6 corresponds to an
experiment. The horizontal axis corresponds to the measurement time for one
data point. The vertical axis corresponds to the minimal absolute value $%
\left\vert g_{a\gamma \gamma }^{\min }\right\vert $ of the coupling $%
g_{a\gamma \gamma }$ which can be measured with a setup, reduced by the value $\left\vert g_{a\gamma \gamma }^{KSVZ}\right\vert $ expected in the
framework of the KSVZ model \cite{axionrate:17}. The horizontal red dashed lines indicate
the values $\left\vert g_{a\gamma \gamma }^{KSVZ}\right\vert =0.97$ and $%
\left\vert g_{a\gamma \gamma }^{DFSZ}\right\vert =0.36$ expected with the KSVZ and
DFSZ models \cite{axionrate:17}. We show as a vertical black dashed line the
expected coherence time $\varkappa _{ax}^{-1}$ of the axion field. In Figure
6, our setup is named DETOX, for "DETector Of aXions with a phase resolved
protocol". We have indicated with blue points the value of $\left\vert g_{a\gamma \gamma }^{\min
}\right\vert $ for the three integration times $T_{\min }$=$2.4$\textit{~}$%
\mathrm{\mu s}$, $T_{\text{opt}}$=$0.2$\textit{~}$\mathrm{ms}$ and $T_{\max
} $=$0.4$\textit{~}$\mathrm{s}$, obtained with the parameters used in
Section \ref{outputSignal}. We find $\left\vert g_{a\gamma \gamma }^{\min
}/g_{a\gamma \gamma }^{KSVZ}\right\vert =1$ for $T=T_{\min }$ and $T=T_{\max
}$ because we have optimized the parameters of the setup to reach this
limit. For $T=T_{\text{opt}}$, $\left\vert g_{a\gamma \gamma }^{\min
}\right\vert $ is smaller because the signal to noise ratio is higher. In
fact, at this working point, we find that $g_{a\gamma \gamma }^{\min }$
becomes smaller than both $g_{a\gamma \gamma }^{KSVZ}$ and $g_{a\gamma
\gamma }^{DFSZ}$. So far, the regime $\left\vert g_{a\gamma \gamma }^{\min
}/g_{a\gamma \gamma }^{DFSZ}\right\vert \sim 1$ has been reached only by the
ADMX-2018 experiment\cite{Du:2018}. Another advantage of our phase resolved
haloscope is that it should enable a real-time detection of possible axions.
Such a feature has been inaccessible to any existing platform so far. The
phase-resolved haloscope should yield a direct access to $\varkappa _{ax}$,
which does not seem obvious with conventional haloscopes (see SI). 

Finally, it is interesting to point some potential limitations of the frequency window which does not enter in the figure of merit metrics. The frequency window for our detection scheme is linked to the magnetic field resilience of the superconducting circuit inducing the cross-Kerr. The gyromagnetic mode has a 'form factor' which is reflected in the coupling $\Omega_{ax}$. We see in figure S1 that it increases with the magnetic field and therefore with the probed mass. The detection is therefore more favorable at high frequencies. Interestingly, in a recent 'path finder' experimental setup \cite{FruyThery:25b}, we have been able to observe that with granular aluminium devices could induce anharmonicity up to 2T and that  frequency conversion could work up to 23 GHz \cite{Thery:24b,FruyThery:25}. This shows that detection windows of up to 20 GHz in the range of 40-60 GHz are not too far from the present experimental context.

\textbf{Conclusion}

As a conclusion, we have proposed a haloscope which exploits the number-number coupling between a readout mode and a gyromagnetic mode in a
microwave cavity, to perform a phase-resolved measurement sensitive to the axion amplitude and phase. Our coupling scheme implies a figure of merit proportional to $g_{a\gamma \gamma }^{2}${} instead of $g_{a\gamma \gamma }^{4}$ as in a
conventional haloscope. Given the smallness of $g_{a\gamma \gamma }$, this
represents a major advantage for axion search. With the realistic set of
parameters assumed in the previous sections, we predict a figure of merit
which exceeds by more than 4 orders of magnitude that of existing detectors.
This should enable the real-time search for axion dark matter. Another
notable step forward is made possible by the resonant conditions required by
our phase-resolved haloscope, $\omega _{1}\approx \tilde{\omega}_{a}$ and $%
\omega _{m}\approx \omega _{2}\approx \omega _{ax}$, instead of $\tilde{%
\omega}_{a}\approx \omega _{1}\approx \omega _{ax}$ for a conventional
haloscope. The tunability of the gyromagnetic mode thereby offers a great
advantage in terms of frequency scanning range. With such expected
performance, a comprehensive test of the axion paradigm should be within
reach.

Beyond the setup of a nonlinear cavity coupled to a gyromagnetic mode
discussed in this work, the measurement principle that we describe opens
further perspectives of experiments. First, the figure of merit $F_{phase}$
of the phase-resolved haloscope compares favorably with respect to $F_{halo}$
in many ways even in conditions where $\omega _{m}$ would not be tunable. In
fact, our proposal can be transposed straightforwardly to a system of two
nonlinearly coupled standard microwave cavities. This would already
represent an enhancement for axion detection even if standard microwave
cavities have a limited tunability in their resonance frequency. Second, the
principle of our detector could also be transposed to the detection of other
weak signals, provided they can be coupled transversely to a set of two
cavities coupled with a $K_{am}$ term (see the term in $\Omega _{ax}$ in Eq.(%
\ref{1})). In particular, our proposal opens new avenues for quantum sensing
of other coherent cosmological radiation such as dark photons, high
frequency gravitational waves or galactic masers, using similar principles.\\

\textbf{Acknowledgements}

We thank P. Fayet, I. Irastorza, Z. Leghtas, R. T. D'Agonolo, W. Legrand, and K.Petraki for
fruitful discussions. We gratefully acknowledge the Hybrid Quantum Circuits group of ENS/ESPCI and, in particular, Jeanne Bally, Matthieu Delbecq, Chlo\'{e} Fruy, Benjamin Hue and Arnaud Th\'{e}ry. We are indebted to M. Brune for many fruitful discussions and, in particular, an illuminating discussion
on measurement induced dephasing. We also gratefully acknowledge the RADES collaboration. This work is supported by the QRADES Quantera project and the
DarkQuantum ERC project. We dedicate this work to the memory of our colleague Prof. Nick Kaiser who helped us a lot throughout this project.\\

\end{document}